\newif\ifpublic\publicfalse
\publictrue

\documentclass[11pt,a4paper]{article}

\usepackage[a4paper,text={160mm,247mm},centering]{geometry}
\usepackage{layout}

\usepackage{setspace}
\usepackage{multicol}
\usepackage[utf8]{inputenc}

\usepackage[english]{babel}
\usepackage{hyphenat}
\usepackage[nosort]{cite}

\newcommand{\rcite}[2][]{%
  ref.~%
  \ifx\relax#1\relax
    \cite{#2}%
  \else
    \cite[#1]{#2}%
  \fi
}
\newcommand{\rcites}[1]{refs.~\cite{#1}}

\usepackage{amsmath, amssymb, amsfonts, amsthm}
\usepackage{mathtools}
\usepackage[normalem]{ulem}

\theoremstyle{plain}

\newtheorem*{theorem*}{Theorem}

\usepackage{mathfixs}
\ProvideMathFix{autobold}
\ProvideMathFix{multskip}
\ProvideMathFix{frac,root,rootclose}

\allowdisplaybreaks[1]

\makeatletter
\def\mr@ignsp#1 {\ifx\:#1\@empty\else #1\expandafter\mr@ignsp\fi}%
\newcommand{\multiref}[1]{\begingroup
\xdef\mr@no@sparg{\expandafter\mr@ignsp#1 \: }%
\def\mr@comma{}%
\@for\mr@refs:=\mr@no@sparg\do{\mr@comma\def\mr@comma{,}\ref{\mr@refs}}%
\endgroup}
\renewcommand{\eqref}[1]{(\multiref{#1})}
\makeatother

\makeatletter
\newcommand{\namedref}[2]{\hyperref[#2]{#1~\ref*{#2}}}%
\newcommand{\namedreff}[2]{\hyperref[#2]{#1\,\ref*{#2}}}%
\newcommand{\secref}{\namedreff{\S}}
\newcommand{\appref}{\namedref{Appendix}}

\newcommand{\figref}{\namedref{Figure}}

\makeatother

\numberwithin{equation}{section}

\newcommand{\eqn}[1]{eq.~\eqref{#1}}

\newcommand{\eqns}[2]{eqs.~\eqref{#1} and~\eqref{#2}}

\providecommand{\href}[2]{#2}

\usepackage[font=small,labelfont=bf]{caption}

\usepackage{graphbox}

\usepackage[compile=false,fonts]{mpostinl}
\begin{mposttex}[]
\usepackage{amsmath,amssymb,amsfonts}
\end{mposttex}
\begin{mpostdef}
	input metaobj;
	
	def drawzigzag(expr pzz)=
	pair zz[];
	zz[1]:= point 0 of pzz ; zz[2]:= point 1 of pzz;
	nczigzag(zz[1])(zz[2]) "coilwidth(0.15xu)", "linewidth(1pt)", "linearc(.01xu)", "arrows(-)", "coilarmA(0)", "coilarmB(0)";
	enddef;
\end{mpostdef}

\begin{mpostdef}
	pair vpos[];
	pair epos[];
	pair ext[];
	pair exta[];
	path paths[];
	picture pic;
	picture pic[];
	picture savepic;
	path cpath;
	path hpath;
	xu:=1cm;
	yu:=1cm;
\end{mpostdef}

\begin{mpostdef}
	def pensize(expr s)=withpen pencircle scaled s enddef;
	def fillshape(expr p,ci,tb,cb)=
	fill p withcolor ci;
	draw p pensize(tb) withcolor cb;
	enddef;
	def filldot(expr z,s,ci)=
	fillshape(fullcircle scaled s shifted z, ci, 0.5pt, 0.0white);
	enddef;
	def filltrig(expr z,s,r,ci)=
	fillshape(((dir 60)--(dir 180)--(dir 300)--cycle) scaled s rotated r shifted z, ci, 0.5pt, 0.0white);
	enddef;
	def fillsqr(expr z,s,r,ci)=
	fillshape(((dir 45)--(dir 135)--(dir 225)--(dir 315)--cycle) scaled s rotated r shifted z, ci, 0.5pt, 0.0white);
	enddef;
	def drawcross(expr z,s,r,t,c)=
	draw ((-0.5,-0.5)--(+0.5,+0.5)) scaled s rotated r shifted z pensize(t) withcolor c;
	draw ((+0.5,-0.5)--(-0.5,+0.5)) scaled s rotated r shifted z pensize(t) withcolor c;
	enddef;
	
	def midarrow (expr p, t) =
	fill (arrowhead subpath(0,arctime(arclength(subpath (0,t) of p)+0.5ahlength) of p) of p) withcolor ecolor;
	enddef;
	
	def midarrowred (expr p, t) =
	fill (arrowhead subpath(0,arctime(arclength(subpath (0,t) of p)+0.5ahlength) of p) of p) withcolor ecolor;
	enddef;
	
	def midarrowreddark (expr p, t) =
	fill (arrowhead subpath(0,arctime(arclength(subpath (0,t) of p)+0.5ahlength) of p) of p) withcolor ecolordark;
	enddef;
	
	def midarrowblue (expr p, t) =
	fill (arrowhead subpath(0,arctime(arclength(subpath (0,t) of p)+0.5ahlength) of p) of p) withcolor bcolor;
	enddef;
	
	def midarrowbluedark (expr p, t) =
	fill (arrowhead subpath(0,arctime(arclength(subpath (0,t) of p)+0.5ahlength) of p) of p) withcolor bcolordark;
	enddef;	
	
	def midarrowgreen (expr p, t) =
	fill (arrowhead subpath(0,arctime(arclength(subpath (0,t) of p)+0.5ahlength) of p) of p) withcolor gcolor;
	enddef;
	
	def midarrowgreenInv (expr p, t) =
	fill (arrowhead subpath(arctime(arclength(subpath (0,t) of p)+0.5ahlength) of p,0) of p) withcolor gcolor;
	enddef;
	
	def dprop expr p=draw p pensize(3pt) withcolor feyncolor enddef;
	def dpropf expr p=draw p pensize(3pt) withcolor feyncolorfaint enddef;
	
	def drawprop expr p=draw p pensize(1.5pt) enddef;
	def drawline expr p=draw p pensize(0.5pt) enddef;
	def drawpos expr p=drawcross(p, 5pt, 0, 1.0pt, 0.0white) enddef;
	def drawvertex expr p=filldot(p, 5pt, 0.5white+0.5red) enddef;
	def drawblackdotscale expr p=filldot(p, 0.4*xu, black) enddef;
	def drawreddotscale expr p=filldot(p, 0.4*xu, red) enddef;
	def drawwhitedotscale expr p=filldot(p, 0.4*xu, white) enddef;
	def drawblackdot expr p=filldot(p, 8pt, black) enddef;
	def drawwhitedot expr p=filldot(p, 8pt, white) enddef;
	def drawlinearrow expr p=draw p pensize(0.5pt); midarrow (p,0.5); enddef;
	def drawproparrow expr p=draw p pensize(1.5pt); midarrow (p,0.5); enddef;
	def drawproparrowdouble expr p=draw p pensize(1.5pt); midarrow (p,0.65); midarrow (reverse p,0.65); enddef;
	def drawwhitedotmr (expr p, sz)=filldot(p, sz*xu, white) enddef;
	def drawblackdotmr (expr p, sz)=filldot(p, sz*xu, black) enddef;
	
	def framed(expr p)=fill bbox p withcolor white; draw bbox p pensize(0.8pt); draw p; enddef;
	def framedd(expr p)=fill bbox p withcolor 0.94white; draw bbox p pensize(1pt); draw p; enddef;
\end{mpostdef}
\begin{mpostdef}
	color ecolor; ecolor:=(.72,.03,.06);
	color bcolor; bcolor:=(.07,.05,.66);
	color gcolor; gcolor:=(.0,.44,.0);
	color ecolordark; ecolordark:=.3[ecolor,black];
	color bcolordark; bcolordark:=.3[bcolor,black];
	color feyncolor; feyncolor:=(.06,.4,.01);
	color feyncolorfaint; feyncolorfaint:=(.12,.8,.02);
	color arrowcolor; arrowcolor:=(0,0,0);
	color faint; faint:=(.7,0.7,0.7);
	color fgray; fgray:=(.7,0.7,0.7);
	vsize := 0.16xu;
	rad:= 2xu;
	cpath:= for i=0 upto 35: 
	dir (i*10) scaled rad ..
	endfor cycle;
	def vert(expr pos)=
	fill fullcircle scaled vsize shifted point pos of cpath withcolor white;
	draw halfcircle scaled vsize rotated (90+10*pos) shifted point pos of cpath withcolor ecolor pensize(1.3pt);
	enddef;
	def vertthick(expr pos)=
	fill fullcircle scaled vsize shifted point pos of cpath withcolor white;
	draw halfcircle scaled vsize rotated (90+10*pos) shifted point pos of cpath withcolor ecolor pensize(2.0pt);
	enddef;
	def vertcol(expr pos,col)=
	fill fullcircle scaled vsize shifted point pos of cpath withcolor white;
	draw halfcircle scaled vsize rotated (90+10*pos) shifted point pos of cpath withcolor col pensize(1.3pt);
	enddef;
	def drawbdry(expr startpos,endpos)=
	draw subpath(startpos,endpos) of cpath withcolor ecolor pensize(1.3pt);
	enddef;
	def drawbdryblack(expr startpos,endpos)=
	draw subpath(startpos,endpos) of cpath withcolor black pensize(1.3pt);
	enddef;
	def drawbdrythick(expr startpos,endpos)=
	draw subpath(startpos,endpos) of cpath withcolor ecolor pensize(2.0pt);
	enddef;
	def drawbdrythickblack(expr startpos,endpos)=
	draw subpath(startpos,endpos) of cpath withcolor black pensize(2.0pt);
	enddef;
	def drawbdrydashed(expr startpos,endpos)=
	draw subpath(startpos,endpos) of cpath withcolor ecolor dashed withdots scaled 0.5 pensize(1.3pt);
	enddef;
	def drawbdrydashedthick(expr startpos,endpos)=
	draw subpath(startpos,endpos) of cpath withcolor ecolor dashed withdots scaled 0.5 pensize(2.0pt);
	enddef;
	def vlabel(expr pos, vscal, tt)=
	vert(pos);
	label(tt,point pos of cpath scaled vscal);
	enddef;
	
	def vlabelcol(expr pos, vscal, tcol, vcol, tt)=
	vertcol(pos, vcol);
	label(tt,point pos of cpath scaled vscal) withcolor tcol;
	enddef;
	
	def ppath(expr p,spos,epos)=
	subpath(arctime(spos*arclength(p))of p,arctime(epos*arclength(p)) of p) of p
	enddef;
\end{mpostdef}

\begin{mpostdef}
	picture nicearrow; currentpicture:= nullpicture;
	ahlengthsave:=ahlength; ahlength:=12pt;
	drawarrow (0xu,0xu){dir 20}..(2xu,0xu) pensize(3pt) withcolor arrowcolor;
	ahlength:=ahlengthsave;
	nicearrow:= currentpicture; currentpicture:= nullpicture;
\end{mpostdef}

\begin{mpostdef}
	eNum:= 2.718281828459045235;
	numeric startangle, lsdist, scalefactor, stepangle;
	path ls;
	
	vardef logspiralder(expr lspos, endpoint, maxangle, stepnumber, enddir) =
	startangle := ((angle(endpoint-lspos) - (maxangle mod 360)) mod 360); 
	stepangle := (1.0maxangle)/(1.0stepnumber);
	lsdist := length(endpoint-lspos); 
	scalefactor := mlog(lsdist+1.0)/(1.0*maxangle)/256;
	ls:=lspos for t = startangle step stepangle until (maxangle + startangle + 0.02): .. (((mexp(256((t-startangle)*scalefactor))-1.0) * cosd(t),(mexp(256((t-startangle)*scalefactor))-1.0)  * sind(t))+lspos) endfor;
	ls:= subpath(1,(length ls) - 1) of ls;
	ls:= ls .. {dir enddir}endpoint;
	ls
	enddef;
\end{mpostdef}

\begin{mpostdef}
def roundedrectangle(expr xa, ya, xb, yb, r) =
  (xa + r, ya) 
  -- (xb - r, ya) {dir 0} .. {dir 90} (xb, ya + r) 
  -- (xb, yb - r) {dir 90} .. {dir 180} (xb - r, yb) 
  -- (xa + r, yb) {dir 180} .. {dir 270} (xa, yb - r) 
  -- (xa, ya + r) {dir 270} .. {dir 0} cycle
enddef;

def SchottkyBackground(expr xoffset, yoffset, w, h) =
fregion := 0.2xu;
fsteps := 80;

for ii = 0 upto fsteps:
  numeric margin, g;
  margin := ii / fsteps * fregion;
  gcol := 1.0 - 0.3 * (ii / fsteps); 

  draw roundedrectangle(margin+xoffset, margin+yoffset,w - margin+xoffset, h-margin+yoffset, fregion * (1-ii / fsteps) ) withcolor (gcol, gcol, gcol) pensize( fregion / (fsteps - 1));
endfor
  fill roundedrectangle(margin+xoffset, margin+yoffset,w - margin+xoffset, h-margin+yoffset, 0 ) withcolor (0.7,0.7,0.7);
enddef;

def SchottkyAxes(expr xstartpos, xendpos, ystartpos, yendpos) = 
  drawarrow xstartpos-- xendpos pensize (1.0pt);  
  drawarrow ystartpos -- yendpos pensize (1.0pt);  
enddef;

def SchottkyCircle(expr centerPos, rad, col, bcol) = 
  fill fullcircle scaled (2*rad) shifted centerPos withcolor bcol; 
  draw fullcircle scaled (2*rad) shifted centerPos withcolor col pensize(0.7pt);
enddef;

def SchottkyCircleNoFilling(expr centerPos, rad, col) = 
  draw fullcircle scaled (2*rad) shifted centerPos withcolor col pensize(0.7pt);
enddef;

def SchottkyACircle(expr centerPos, rad, col, bcol) = 
  fill fullcircle scaled (2*rad) shifted centerPos withcolor bcol;
  drawarrow reverse(fullcircle) scaled (2*rad) shifted centerPos withcolor col pensize(0.7pt);
enddef;

def SchottkyCircleDashed(expr centerPos, rad, col, bcol) = 
	fill fullcircle scaled (2*rad) shifted centerPos withcolor bcol;
	draw fullcircle scaled (2*rad) shifted centerPos withcolor col pensize(0.7pt) dashed evenly;
enddef;

def SchottkyCircleDashedNoFilling(expr centerPos, rad, col) = 
	draw fullcircle scaled (2*rad) shifted centerPos withcolor col pensize(0.7pt) dashed evenly;
enddef;

def SchottkyCross(expr PosA, Sangle, col) = 
  transform ST;
  ST := identity rotated (Sangle + 45 + 15) shifted PosA; 
  draw ((-0.05xu,0xu)--(0.05xu,0xu)) transformed ST withcolor col;
  draw ((0,-0.05xu)--(0,0.05xu)) transformed ST withcolor col;
enddef;

def SchottkyCrossS(expr PosA, Sangle, size, col) = 
  transform ST;
  ST := identity rotated (Sangle + 45 + 15) shifted PosA; 
  draw ((-size,0xu)--(size,0xu)) transformed ST withcolor col;
  draw ((0,-size)--(0,size)) transformed ST withcolor col;
enddef;

def SchottkySquare(expr PosA, Sangle) = 
	transform ST;
	ST := identity rotated (Sangle + 45 + 15) shifted PosA; 
	draw ((-0.03xu,-0.03xu)--(0.03xu,-0.03xu)--(0.03xu,0.03xu)--(-0.03xu,0.03xu)--(-0.03xu,-0.03xu)) transformed ST;
enddef;

def SchottkyPathArrow(expr PosA,PosB) =
  path SPA;
  numeric SPAdir;
  SPAdir := angle(PosB-PosA);
  SPA := PosA {dir (SPAdir-15)} .. {dir (SPAdir+15)} PosB;
  drawarrow subpath (0.04 * length SPA, 0.96 * length SPA) of SPA pensize(0.7pt);
  SchottkyCross (PosB, SPAdir, (0,0,0));
enddef;

def SchottkyPathArrowAlt(expr PosA,PosB) =
  path SPA;
  numeric SPAdir;
  SPAdir := angle(PosB-PosA);
  SPA := PosA {dir (SPAdir+15)} .. {dir (SPAdir-15)} PosB;
  drawarrow subpath (0.04 * length SPA, 0.96 * length SPA) of SPA pensize(0.7pt);
  SchottkyCross (PosB, SPAdir, (0,0,0));
enddef;
\end{mpostdef}

\begin{mpostdef}
	path quartercircle;
	quartercircle := subpath (0, 2) of fullcircle;
\end{mpostdef}

\begin{mpostfig}[label=associatorrelation]
	Retau := 1.3xu;
	Imtau := 4.0xu;
	maxX := 6.0xu;
	maxY := 4.5xu;
	markedPointX := 2.7xu;
	markedPointY := 1.7xu;
	
	ahlength := 4bp;
	drawarrow (-0.1xu, 0xu) -- (maxX, 0xu) pensize (1pt);  
	drawarrow (0xu, -0.1xu) -- (0xu, maxY) pensize (1pt);  
	label.bot(btex $\textrm{Re}(z)$ etex, (maxX-0.5xu, 0xu));
	label.lft(btex $\textrm{Im}(z)$ etex, (0xu, maxY-0.5xu));
	
	ahlength := 0.2cm;
	draw (Retau,Imtau) -- (maxX-0.5xu,Imtau) pensize (1pt) withcolor ecolor;
	midarrowred((maxX-0.5xu,Imtau)--(Retau,Imtau),0.5);
	draw (0.0xu,0.0xu) -- (maxX-Retau-0.5xu,0.0xu) pensize (1pt) withcolor ecolor;
	midarrowred((0.0xu,0.0xu)--(maxX-Retau-0.5xu,0.0xu),0.5);
	draw (0.0xu,0.0xu) -- (Retau,Imtau) pensize (1pt) withcolor bcolor;
	midarrowblue((Retau,Imtau)--(0.0xu,0.0xu),0.5);
	draw (maxX-0.5xu,Imtau) -- (maxX-Retau-0.5xu,0.0xu) pensize (1pt) withcolor bcolor;
	midarrowblue((maxX-Retau-0.5xu,0.0xu)--(maxX-0.5xu,Imtau),0.5);
	
	label.top(btex $\tau$ etex, (Retau,Imtau));
	label.bot(btex $1$ etex, (maxX-Retau-0.5xu, 0.0xu));
	label.top(btex $1+\tau$ etex, (maxX-0.5xu, Imtau));
	
	label.bot(btex $\mathfrak{A}$ etex, (0.5 * (maxX-Retau-0.5xu),0.0xu));
	label.rt(btex $\mathfrak{B}$ etex, (maxX-0.5*Retau-0.5xu,0.5*Imtau));
	label.top(btex $\mathfrak{A}^{-1}$ etex, (0.5 * (maxX+Retau-0.5xu),Imtau));
	label.rt(btex $\mathfrak{B}^{-1}$ etex, (0.5*Retau,0.5*Imtau));
	
	dotlabeldiam:=4pt;
	dotlabel.top (btex $ $ etex, (markedPointX,markedPointY));

	ahlength := 0.2cm;
	cycleRadius= 1.0xu;
	draw fullcircle xscaled (cycleRadius) yscaled (cycleRadius) shifted (markedPointX,markedPointY) pensize(1pt) withcolor gcolor;	
	midarrowgreen(fullcircle xscaled (cycleRadius) yscaled (cycleRadius) shifted (markedPointX,markedPointY),0.5);
	label.rt(btex $\mathfrak{C}$ etex, (markedPointX+0.5xu,markedPointY+0.2xu));
	pic[1]:=currentpicture;
	currentpicture:=nullpicture;
	
	draw fullcircle xscaled 8xu yscaled 3.8xu pensize (1pt);
	draw (-2xu,0.06xu){dir -20}..(2xu,0.06xu) pensize (1.0pt);
	draw (-1.6xu,-0.04xu){dir 25}..(1.6xu,-0.04xu) pensize (1.0pt);
	path apath; apath = (0xu,-0.3xu) .. controls (-0.2xu,-0.36xu) and (-0.2xu,-1.84xu) .. ( 0xu,-1.9xu);
	path bpath; bpath = (-4xu,0xu) ..controls (-4xu,-1.5xu) and (4xu,-1.5xu) .. (4xu,0xu);
	draw (-4xu,0xu) ..controls (-4xu,1.5xu) and (4xu,1.5xu) .. ( 4xu,0xu) dashed evenly pensize (1.pt) withcolor bcolor;
	draw (0xu,-0.3xu) .. controls (0.2xu,-0.36xu) and (0.2xu,-1.84xu) .. ( 0xu,-1.9xu) dashed evenly pensize (1.pt) withcolor ecolor;
	draw fullcircle xscaled 8xu yscaled 3.8xu pensize (1pt);
	draw (-2xu,0.06xu){dir -20}..(2xu,0.06xu) pensize (1.0pt);
	draw bpath pensize (1.pt) withcolor bcolor;
	draw apath pensize (1.pt) withcolor ecolor;
	dotlabeldiam:=4pt;
	dotlabel.bot (btex $ $ etex, (1.7xu,-0.6xu));
	path cpath; cpath = fullcircle xscaled 0.7xu yscaled 0.6xu rotated (10) shifted (1.69xu,-0.6xu);
	draw cpath withcolor gcolor pensize (1pt);
	ahlength := 0.2cm;
	midarrowblue(bpath,0.4);
	midarrowred(apath,0.4);
	midarrowgreen(cpath,0.5);
	label.lft(btex $\mathfrak{A}$ etex, point 0.4 of apath);
	label.top(btex $\mathfrak{B}$ etex, point 0.4 of bpath);
	label.rt(btex $\mathfrak{C}$ etex, point 0.4 of cpath);
	pic[2]:=currentpicture;
	currentpicture:=nullpicture;
	
	draw pic[1] shifted (0xu,0xu);
	draw pic[2] shifted (10.8xu,2xu);
\end{mpostfig}

\begin{mpostfig}[label=ellipticassociators]
	Retau := 1.3xu;
	Imtau := 4.0xu;
	maxX := 6.0xu;
	maxY := 4.5xu;
	markedPointX := 2.7xu;
	markedPointY := 1.7xu;
	
	ahlength := 4bp;
	drawarrow (-0.1xu, 0xu) -- (maxX, 0xu) pensize (1pt);  
	drawarrow (0xu, -0.1xu) -- (0xu, maxY) pensize (1pt);  
	label.bot(btex $\textrm{Re}(z)$ etex, (maxX-0.5xu, 0xu));
	label.lft(btex $\textrm{Im}(z)$ etex, (0xu, maxY-0.5xu));
	
	ahlength := 0.2cm;
	phaseRadius := 0.5xu;
	associatorPenSize := 1.7pt;
	draw (Retau,Imtau) -- (maxX-0.5xu,Imtau) pensize (1pt) withcolor ecolor;
	draw (0.0xu,0.0xu) -- (maxX-Retau-0.5xu,0.0xu) pensize (1pt) withcolor ecolor;
	draw (0.5*phaseRadius,0.0xu) -- (maxX-Retau-0.5xu-0.5*phaseRadius,0.0xu) pensize (associatorPenSize) withcolor ecolor;
	midarrowred((0.0xu,0.0xu)--(maxX-Retau-0.5xu,0.0xu),0.5);
	draw (0.0xu,0.0xu) -- (Retau,Imtau) pensize (1pt) withcolor bcolor;
	eps := 0.5*phaseRadius/sqrt(Retau**2+Imtau**2);
	draw (maxX-0.5xu,Imtau) -- (maxX-Retau-0.5xu,0.0xu) pensize (1pt) withcolor bcolor;
	draw (eps*4.2xu,0.0xu) -- (Retau+eps*4.2xu,Imtau) pensize (associatorPenSize) withcolor bcolor;
	draw halfcircle scaled phaseRadius shifted (maxX-Retau-0.5xu,0.0xu) pensize(associatorPenSize) withcolor ecolor;
	midarrowblue((eps*4.2xu,0.0xu)--(Retau+eps*4.2xu,Imtau),0.52);
	
	label.lft(btex $\tau$ etex, (Retau,Imtau));
	label.bot(btex $1$ etex, (maxX-Retau-0.5xu, 0.0xu));
	label.top(btex $1+\tau$ etex, (maxX-0.5xu, Imtau));
	
	label.bot(btex $A(\tau)$ etex, (0.5 * (maxX-Retau-0.5xu),0.0xu));
	label.rt(btex $B(\tau)$ etex, (0.5*Retau+eps*4.2xu,0.5*Imtau));
	pic[1]:=currentpicture;
	currentpicture:=nullpicture;
	
	draw fullcircle xscaled 8xu yscaled 3.8xu pensize (1pt);
	draw (-2xu,0.06xu){dir -20}..(2xu,0.06xu) pensize (1.0pt);
	draw (-1.6xu,-0.04xu){dir 25}..(1.6xu,-0.04xu) pensize (1.0pt);
	draw (-4xu,0xu) ..controls (-4xu,1.5xu) and (4xu,1.5xu) .. ( 4xu,0xu) dashed evenly pensize (0.75pt) withcolor bcolor;
	draw (0xu,-0.3xu) .. controls (0.2xu,-0.36xu) and (0.2xu,-1.84xu) .. ( 0xu,-1.9xu) dashed evenly pensize (0.75pt) withcolor ecolor;
	draw (-4xu,0xu) ..controls (-4xu,-1.5xu) and (4xu,-1.5xu) .. (4xu,0xu) pensize (0.75pt) withcolor bcolor;
	draw (0xu,-0.3xu) .. controls (-0.2xu,-0.36xu) and (-0.2xu,-1.84xu) .. ( 0xu,-1.9xu) pensize (0.75pt) withcolor ecolor;
	xwidth := 4xu;
	draw (-xwidth,0.0xu) ..controls (-xwidth,1.5xu) and (xwidth,1.5xu) .. (xwidth,0.0xu) dashed evenly pensize (1.5pt) withcolor bcolor;
	draw (0.0xu,-0.3xu) .. controls (0.2xu,-0.36xu) and (0.2xu,-1.84xu) .. (0.0xu,-1.9xu) dashed evenly pensize (1.5pt) withcolor ecolor;
	draw (-2xu,0.06xu){dir -20}..(2xu,0.06xu) pensize (1.0pt);
	draw (-4xu,0xu) ..controls (-4xu,-1.5xu) and (4xu,-1.5xu) .. (4xu,0xu) pensize (0.75pt) withcolor bcolor;
	draw fullcircle xscaled 8xu yscaled 3.8xu pensize (1pt);
	path bpath; bpath = (-xwidth,0.0xu) ..controls (-xwidth,-1.5xu) and (xwidth,-1.5xu) .. (xwidth,0.0xu);
	draw subpath (0,.4555) of bpath pensize (1.5pt) withcolor bcolor;
	midarrowblue(bpath,0.3);
	draw subpath (0.535,1) of bpath pensize (1.5pt) withcolor bcolor;
	draw subpath (0,-4.0) of fullcircle scaled 0.77xu shifted (-0.15xu,-1.12xu) pensize (1.5pt) withcolor bcolor dashed withdots scaled 0.4;
	path rpath; rpath = (0.0xu,-0.3xu) .. controls (-0.2xu,-0.36xu) and (-0.2xu,-1.84xu) .. (0.0xu,-1.9xu);
	draw subpath (0,0.41) of rpath pensize (1.5pt) withcolor ecolor;
	draw subpath (0.73,1) of rpath pensize (1.5pt) withcolor ecolor;
	midarrowred(rpath,0.3);
	draw subpath (-1.95,1.95) of fullcircle scaled 0.43xu shifted (-0.15xu,-1.12xu) pensize (1.5pt) withcolor ecolor;
	dotlabeldiam:=4pt;
	dotlabel.bot (btex $ $ etex, point .51 of ((0xu,-0.3xu) .. controls (-0.2xu,-0.36xu) and (-0.2xu,-1.84xu) .. ( 0xu,-1.9xu)));
	pic[2]:=currentpicture;
	currentpicture:=nullpicture;
	
	draw pic[1] shifted (0xu,0xu);
	draw pic[2] shifted (10.8xu,2xu);
\end{mpostfig}

\begin{mpostfig}[label=ellipticsvcondition]
	Retau := 1.3xu;
	Imtau := 4.0xu;
	maxX := 6.0xu;
	maxY := 4.5xu;
	markedPointX := 2.7xu;
	markedPointY := 1.7xu;
	
	ahlength := 4bp;
	drawarrow (-0.1xu, 0xu) -- (maxX, 0xu) pensize (1pt);  
	drawarrow (0xu, -0.1xu) -- (0xu, maxY) pensize (1pt);  
	label.bot(btex $\textrm{Re}(z)$ etex, (maxX-0.5xu, 0xu));
	label.lft(btex $\textrm{Im}(z)$ etex, (0xu, maxY-0.5xu));
	
	ahlength := 0.2cm;
	phaseRadius := 0.5xu;
	associatorPenSize := 1.7pt;
	draw (Retau,Imtau) -- (maxX-0.5xu,Imtau) pensize (1pt) withcolor ecolor;
	draw (0.0xu,0.0xu) -- (maxX-Retau-0.5xu,0.0xu) pensize (1pt) withcolor ecolor;
	draw (0.0xu,0.0xu) -- (Retau,Imtau) pensize (1pt) withcolor bcolor;
	draw (maxX-0.5xu,Imtau) -- (maxX-Retau-0.5xu,0.0xu) pensize (1pt) withcolor bcolor;
	
	draw (0.5*phaseRadius,0.1xu) -- (maxX-Retau-0.5xu-0.5*phaseRadius,0.1xu) pensize (associatorPenSize) withcolor ecolor;
	draw subpath (8,11.45) of fullcircle scaled phaseRadius shifted (maxX-Retau-0.5xu,0.0xu) pensize(associatorPenSize) withcolor ecolor;
	draw subpath (4.55,8) of fullcircle scaled phaseRadius shifted (maxX-Retau-0.5xu,0.0xu) pensize(associatorPenSize) withcolor ecolordark;
	draw (maxX-Retau-0.5xu-0.5*phaseRadius,-0.1xu) -- (0.5*phaseRadius,-0.1xu) pensize (associatorPenSize) withcolor ecolordark;
	midarrowred((0.1xu,0.1xu)--(maxX-Retau-0.5xu,0.1xu),0.5);
	midarrowreddark((maxX-Retau-0.5xu,-0.1xu)--(0.1xu,-0.1xu),0.5);
	eps := 0.5*phaseRadius/sqrt(Retau**2+Imtau**2);
	draw subpath (eps,1) of ((0.15xu,0.0xu) -- (Retau+0.15xu,Imtau)) pensize (associatorPenSize) withcolor bcolor;
	draw subpath (1,eps) of ((-0.15xu,0.0xu) -- (Retau-0.15xu,Imtau)) pensize (associatorPenSize) withcolor bcolordark;
	draw subpath (2.,4.5) of fullcircle scaled 0.3xu shifted (-sqrt(Retau*Retau+Imtau*Imtau),0.0xu) rotated (180+72) pensize (associatorPenSize) withcolor bcolor dashed withdots scaled 0.4;
	draw subpath (4.5,6.5) of fullcircle scaled 0.3xu shifted (-sqrt(Retau*Retau+Imtau*Imtau),0.0xu) rotated (180+72) pensize (associatorPenSize) withcolor bcolordark dashed withdots scaled 0.4;
	midarrowblue((0.15xu,0.0xu)--(Retau+0.15xu,Imtau),0.5);
	midarrowbluedark((Retau-0.15xu,Imtau)--(-0.15xu,0.0xu),0.5);
	
	label.lft(btex $\tau$ etex, (Retau-0.2xu,Imtau));
	label.bot(btex $1$ etex, (maxX-Retau-0.5xu, -0.3xu));
	label.top(btex $1+\tau$ etex, (maxX-0.5xu, Imtau));
	
	draw (0xu, 1.8xu) -- (0xu, 2.2xu) pensize (5pt) withcolor white;
	
	label.top(btex $A_{\mathcal{Y}}$ etex, (0.5 * (maxX-Retau-0.5xu),0.1xu));
	label.bot(btex $\mathrm{R}(\overline{A_{\mathcal{Y}'}})$ etex, (0.5 * (maxX-Retau-0.5xu),-0.1xu));
	label.rt(btex $B_{\mathcal{Y}}$ etex, (0.5*Retau+0.15xu,0.5*Imtau));
	label.lft(btex $\mathrm{R}(\overline{B_{\mathcal{Y}'}})$ etex, (0.5*Retau-0.15xu,0.5*Imtau));
	pic[1]:=currentpicture;
	currentpicture:=nullpicture;
	
	draw fullcircle xscaled 8xu yscaled 3.8xu pensize (1pt);
	draw (-2xu,0.06xu){dir -20}..(2xu,0.06xu) pensize (1.0pt);
	draw (-1.6xu,-0.04xu){dir 25}..(1.6xu,-0.04xu) pensize (1.0pt);
	draw (-4xu,0xu) ..controls (-4xu,1.5xu) and (4xu,1.5xu) .. ( 4xu,0xu) dashed evenly pensize (0.75pt) withcolor bcolor;
	draw (0xu,-0.3xu) .. controls (0.2xu,-0.36xu) and (0.2xu,-1.84xu) .. ( 0xu,-1.9xu) dashed evenly pensize (0.75pt) withcolor ecolor;
	draw (-4xu,0xu) ..controls (-4xu,-1.5xu) and (4xu,-1.5xu) .. (4xu,0xu) pensize (0.75pt) withcolor bcolor;
	draw (0xu,-0.3xu) .. controls (-0.2xu,-0.36xu) and (-0.2xu,-1.84xu) .. ( 0xu,-1.9xu) pensize (0.75pt) withcolor ecolor;
	xwidth := 3.98xu;
	draw subpath (0.03,0.97) of ((-xwidth,0.1xu) ..controls (-xwidth,1.6xu) and (xwidth,1.6xu) .. (xwidth,0.1xu)) dashed evenly pensize (1.5pt) withcolor bcolordark;
	draw subpath (0.03,0.97) of ((-xwidth,-0.1xu) ..controls (-xwidth,1.4xu) and (xwidth,1.4xu) .. (xwidth,-0.1xu)) dashed evenly pensize (1.5pt) withcolor bcolor;
	draw (0.1xu,-0.3xu) .. controls (0.3xu,-0.36xu) and (0.3xu,-1.84xu) .. (0.1xu,-1.9xu) dashed evenly pensize (1.5pt) withcolor ecolor;
	draw (-0.1xu,-0.3xu) .. controls (0.1xu,-0.36xu) and (0.1xu,-1.84xu) .. (-0.1xu,-1.9xu) dashed evenly pensize (1.5pt) withcolor ecolordark;
	draw fullcircle xscaled 8xu yscaled 3.8xu pensize (1pt);
	draw (-2xu,0.06xu){dir -20}..(2xu,0.06xu) pensize (1.0pt);
	path bpathu; bpathu = (-xwidth,0.1xu) ..controls (-xwidth,-1.4xu) and (xwidth,-1.4xu) .. (xwidth,0.1xu);
	draw subpath (0,.459) of bpathu pensize (1.5pt) withcolor bcolordark;
	draw subpath (0.53,1) of bpathu pensize (1.5pt) withcolor bcolordark;
	path bpathl; bpathl = (-xwidth,-0.1xu) ..controls (-xwidth,-1.6xu) and (xwidth,-1.6xu) .. (xwidth,-0.1xu);
	draw subpath (0,.459) of bpathl pensize (1.5pt) withcolor bcolor;
	draw subpath (0.53,1) of bpathl pensize (1.5pt) withcolor bcolor;
	draw subpath (-3.55,0) of fullcircle scaled 0.7xu shifted (-0.15xu,-1.12xu) pensize (1.5pt) withcolor bcolor dashed withdots scaled 0.3;
	draw subpath (0,3.57) of fullcircle scaled 0.7xu shifted (-0.15xu,-1.12xu) pensize (1.5pt) withcolor bcolordark dashed withdots scaled 0.3;
	path rpathr; rpathr = (0.1xu,-0.3xu) .. controls (-0.1xu,-0.36xu) and (-0.1xu,-1.84xu) .. (0.1xu,-1.9xu);
	draw subpath (0,0.41) of rpathr pensize (1.5pt) withcolor ecolor;
	draw subpath (0.72,1) of rpathr pensize (1.5pt) withcolor ecolor;
	draw (-4xu,0xu) ..controls (-4xu,-1.5xu) and (4xu,-1.5xu) .. (4xu,0xu) pensize (0.75pt) withcolor bcolor;
	draw subpath (-5.5,-1.98) of fullcircle scaled 0.5xu shifted (-0.15xu,-1.12xu) pensize (1.5pt) withcolor ecolordark;
	draw subpath (-1.98,1.46) of fullcircle scaled 0.5xu shifted (-0.15xu,-1.12xu) pensize (1.5pt) withcolor ecolor;
	path rpathl; rpathl = (-0.1xu,-0.3xu) .. controls (-0.3xu,-0.36xu) and (-0.3xu,-1.84xu) .. (-0.1xu,-1.9xu);
	draw subpath (0,0.4) of rpathl pensize (1.5pt) withcolor ecolordark;
	draw subpath (0.72,1) of rpathl pensize (1.5pt) withcolor ecolordark;
	dotlabeldiam:=4pt;
	dotlabel.bot (btex $ $ etex, point .51 of ((0xu,-0.3xu) .. controls (-0.2xu,-0.36xu) and (-0.2xu,-1.84xu) .. ( 0xu,-1.9xu)));
	midarrowblue(bpathl,0.4);
	midarrowbluedark(reverse bpathu,0.6);
	midarrowred(rpathr,0.3);
	midarrowreddark(reverse rpathl,0.75);
	pic[2]:=currentpicture;
	currentpicture:=nullpicture;
	
	draw pic[1] shifted (0xu,0xu);
	draw pic[2] shifted (10.8xu,2xu);
\end{mpostfig}

\usepackage{graphicx}
\usepackage{subcaption}
\usepackage{caption}

\usepackage[dvipsnames,table]{xcolor}
\definecolor{mygreen}{rgb}{0,0.4,0}
\definecolor{myblue}{rgb}{0,0.0,0.4}
\definecolor{refrcolor}{rgb}{0,0.4,0}
\definecolor{cgreen}{rgb}{0,0.7,0}
\definecolor{ecolor}{rgb}{.52,.03,.06}
\definecolor{bgcolor}{rgb}{.96,.95,.80}
\definecolor{bgcolordark}{rgb}{.80,.80,.67}
\definecolor{faint}{rgb}{.80,.80,.80}

\usepackage{tcolorbox}
\tcbuselibrary{breakable, skins}
\usetikzlibrary{patterns}

\newtcolorbox{myproofbox}[1][]{
	enhanced,
	breakable,
	borderline west={3pt}{0pt}{faint},
	notitle,
	before skip=10pt,
	after skip=10pt,
	colback=white, 
	colframe=white,
	frame hidden,
	boxrule=0pt, 
	boxsep=0pt,
	sharp corners,
	left=9pt, right=0pt, top=1pt, bottom=0pt,
	fontupper=\small,
	#1
}

\makeatletter
	{\popQED\endtrivlist\end{myproofbox}\@endpefalse%
	\if@noskipsec\leavevmode\fi\noindent\ignorespacesafterend}
\makeatother

\newtheorem{example}[theorem]{Example}

\newtcolorbox{myexamplebox}[1][]{%
	enhanced,
	breakable,
	borderline west={3pt}{0pt}{faint},
	notitle,
	before skip=10pt,
	after skip=10pt,
	colback=white, 
	colframe=white,
	frame hidden,
	boxrule=0pt, 
	boxsep=0pt,
	sharp corners,
	left=9pt, right=0pt, top=1pt, bottom=0pt,
	#1
}

\makeatletter
	{\endtrivlist\end{myexamplebox}\@endpefalse%
	\if@noskipsec\leavevmode\fi\noindent\ignorespacesafterend}
\makeatother

\usepackage{enumitem}

\usepackage{tabularx}
\usepackage{booktabs}

\newcounter{sitcounter}

\newcommand{\itemwithlabel}[3][n]{%
  \def\tempposarg{#1} 
  \def\templeftarg{l}
  \def\temprightarg{r}
  \vspace{1em} 
  \noindent
  \refstepcounter{sitcounter}
  \ifx\tempposarg\templeftarg 
  \begin{tabularx}{\linewidth}{@{}m{10em}@{\hspace{2em}}X@{}}
    \raisebox{\dimexpr\ht\strutbox-\height}{#2} & \textbf{Situation \arabic{sitcounter}:} #3
  \end{tabularx}%
  \else
  \ifx\tempposarg\temprightarg 
  \begin{tabularx}{\linewidth}{@{}X@{\hspace{2em}}m{10em}@{}}
    \textbf{Situation \arabic{sitcounter}:} #3 & \raisebox{\dimexpr\ht\strutbox-\height}{#2}
  \end{tabularx}%
  \else
  \begin{tabularx}{\linewidth}{@{\hspace{2em}}X@{}}
    \textbf{Situation \arabic{sitcounter}:} #3
  \end{tabularx}%
  \fi
  \fi
  \par
}

\newlist{situations}{enumerate}{1}
\setlist[situations]{
    labelsep=0.8em,
    leftmargin=0pt, 
    align=left,
    labelindent=0in, 
    labelwidth=0in,
    itemindent=0.8em,
    label=\textbf{Situation \arabic*:},
    ref=\arabic*
}

\makeatletter
\providecommand*{\shuffle}{%
  \mathbin{\mathpalette\shuffle@{}}%
}
\newcommand*{\shuffle@}[2]{%
  \sbox0{$#1\vcenter{}$}%
  \kern .15\ht0 
  \rlap{\vrule height .25\ht0 depth 0pt width 2.5\ht0}%
  \raise.1\ht0\hbox to 2.5\ht0{%
    \vrule height 1.75\ht0 depth -.1\ht0 width .17\ht0 %
    \hfill
    \vrule height 1.75\ht0 depth -.1\ht0 width .17\ht0 %
    \hfill
    \vrule height 1.75\ht0 depth -.1\ht0 width .17\ht0 %
  }%
  \kern .15\ht0 
}
\makeatother

\usepackage{xparse}

\ExplSyntaxOn
\NewDocumentCommand{\Gtargz}{m m}
{
 \Gt\left(\begin{smallmatrix}
 \Gtargz_print:n {#1} \\
 \Gtargz_print:n {#2}
 \end{smallmatrix};z\right)
}
\seq_new:N \l_Gtargz_list_seq
\cs_new_protected:Npn \Gtargz_print:n #1
{
  \seq_set_split:Nnn \l_Gtargz_list_seq { , } { #1 }
  \seq_use:Nn \l_Gtargz_list_seq { , & }
}
\ExplSyntaxOff

\ExplSyntaxOn
\NewDocumentCommand{\Gtargt}{m m}
{
 \Gt\left(\begin{smallmatrix}
 \Gtargt_print:n {#1} \\
 \Gtargt_print:n {#2}
 \end{smallmatrix};t\right)
}
\seq_new:N \l_Gtargt_list_seq
\cs_new_protected:Npn \Gtargt_print:n #1
{
  \seq_set_split:Nnn \l_Gtargt_list_seq { , } { #1 }
  \seq_use:Nn \l_Gtargt_list_seq { , & }
}
\ExplSyntaxOff

\ExplSyntaxOn
\NewDocumentCommand{\Gtargtheta}{m m}
{
 \Gt\left(\begin{smallmatrix}
 \Gtargtheta_print:n {#1} \\
 \Gtargtheta_print:n {#2}
 \end{smallmatrix};\vt\right)
}
\seq_new:N \l_Gtargtheta_list_seq
\cs_new_protected:Npn \Gtargtheta_print:n #1
{
  \seq_set_split:Nnn \l_Gtargtheta_list_seq { , } { #1 }
  \seq_use:Nn \l_Gtargtheta_list_seq { , & }
}
\ExplSyntaxOff

\ExplSyntaxOn
\NewDocumentCommand{\Gtargzt}{m m}
{
 \Gt\left(\begin{smallmatrix}
 \Gtargzt_print:n {#1} \\
 \Gtargzt_print:n {#2}
 \end{smallmatrix};z|\tau\right)
}
\seq_new:N \l_Gtargzt_list_seq
\cs_new_protected:Npn \Gtargzt_print:n #1
{
  \seq_set_split:Nnn \l_Gtargzt_list_seq { , } { #1 }
  \seq_use:Nn \l_Gtargzt_list_seq { , & }
}
\ExplSyntaxOff

\ExplSyntaxOn
\NewDocumentCommand{\Gtargxit}{m m}
{
 \Gt\left(\begin{smallmatrix}
 \Gtargxit_print:n {#1} \\
 \Gtargxit_print:n {#2}
 \end{smallmatrix};\xi|\tau\right)
}
\seq_new:N \l_Gtargxit_list_seq
\cs_new_protected:Npn \Gtargxit_print:n #1
{
  \seq_set_split:Nnn \l_Gtargxit_list_seq { , } { #1 }
  \seq_use:Nn \l_Gtargxit_list_seq { , & }
}
\ExplSyntaxOff

\ExplSyntaxOn
\NewDocumentCommand{\Gtargtt}{m m}
{
 \Gt\left(\begin{smallmatrix}
 \Gtargtt_print:n {#1} \\
 \Gtargtt_print:n {#2}
 \end{smallmatrix};t|\tau\right)
}
\seq_new:N \l_Gtargtt_list_seq
\cs_new_protected:Npn \Gtargtt_print:n #1
{
  \seq_set_split:Nnn \l_Gtargtt_list_seq { , } { #1 }
  \seq_use:Nn \l_Gtargtt_list_seq { , & }
}
\ExplSyntaxOff

\ExplSyntaxOn
\NewDocumentCommand{\Gtargzg}{m m}
{
 \Gt\left(\begin{smallmatrix}
 \Gtargzg_print:n {#1} \\
 \Gtargzg_print:n {#2}
 \end{smallmatrix};z|\SGroup\right)
}
\seq_new:N \l_Gtargzg_list_seq
\cs_new_protected:Npn \Gtargzg_print:n #1
{
  \seq_set_split:Nnn \l_Gtargzg_list_seq { , } { #1 }
  \seq_use:Nn \l_Gtargzg_list_seq { , & }
}
\ExplSyntaxOff

\ExplSyntaxOn
\NewDocumentCommand{\Gtargxi}{m m}
{
 \Gt\left(\begin{smallmatrix}
 \Gtargxi_print:n {#1} \\
 \Gtargxi_print:n {#2}
 \end{smallmatrix};\xi\right)
}
\seq_new:N \l_Gtargxi_list_seq
\cs_new_protected:Npn \Gtargxi_print:n #1
{
  \seq_set_split:Nnn \l_Gtargxi_list_seq { , } { #1 }
  \seq_use:Nn \l_Gtargxi_list_seq { , & }
}
\ExplSyntaxOff

\ExplSyntaxOn
\NewDocumentCommand{\Gtargxig}{m m}
{
 \Gt\left(\begin{smallmatrix}
 \Gtargxig_print:n {#1} \\
 \Gtargxig_print:n {#2}
 \end{smallmatrix};\xi|\SGroup\right)
}
\seq_new:N \l_Gtargxig_list_seq
\cs_new_protected:Npn \Gtargxig_print:n #1
{
  \seq_set_split:Nnn \l_Gtargxig_list_seq { , } { #1 }
  \seq_use:Nn \l_Gtargxig_list_seq { , & }
}
\ExplSyntaxOff

\ExplSyntaxOn
\NewDocumentCommand{\Gtargtg}{m m}
{
 \Gt\left(\begin{smallmatrix}
 \Gtargtg_print:n {#1} \\
 \Gtargtg_print:n {#2}
 \end{smallmatrix};t|\SGroup\right)
}
\seq_new:N \l_Gtargtg_list_seq
\cs_new_protected:Npn \Gtargtg_print:n #1
{
  \seq_set_split:Nnn \l_Gtargtg_list_seq { , } { #1 }
  \seq_use:Nn \l_Gtargtg_list_seq { , & }
}
\ExplSyntaxOff

\newcommand{\SI}[1]{\Sel[#1]}

\ExplSyntaxOn
\NewDocumentCommand{\SIE}{m m}
{
\SelE\!\Big[\begin{smallmatrix}
 \SI_print:n {#1} \\
 \SI_print:n {#2}
 \end{smallmatrix}\Big]
}
\seq_new:N \l_SI_list_seq
\cs_new_protected:Npn \SI_print:n #1
{
  \seq_set_split:Nnn \l_SI_list_seq { , } { #1 }
  \seq_use:Nn \l_SI_list_seq { , & }
}
\ExplSyntaxOff

\ExplSyntaxOn
\NewDocumentCommand{\Gargbare}{m m m m}
{
	\Gt\left(\begin{smallmatrix}
		\Gargbare_print:n {#1} \\
		\Gargbare_print:n {#2}
	\end{smallmatrix}\,\Big|\,\Gargbare_print:n {#3}\,;\Gargbare_print:n {#4}\right)
}
\seq_new:N \l_Gargbare_list_seq
\cs_new_protected:Npn \Gargbare_print:n #1
{
	\seq_set_split:Nnn \l_Gargbare_list_seq {,} { #1 }
	\seq_use:Nn \l_Gargbare_list_seq {\hspace{-0.1em},\hspace{-0.18em} & }
}
\ExplSyntaxOff

\newcommand{\Greg}{\tilde{\Gamma}_{\text{reg}}}

\ExplSyntaxOn
\NewDocumentCommand{\Gtreg}{m m m m}
{
	\Greg\!\left(\!\begin{smallmatrix}
		\Gtreg_print:n {#1} \\
		\Gtreg_print:n {#2}
	\end{smallmatrix}\!\,\Big|\,\Gtreg_print:n {#3}\,;\Gtreg_print:n {#4}\right)
}
\seq_new:N \l_Gtreg_list_seq
\cs_new_protected:Npn \Gtreg_print:n #1
{
	\seq_set_split:Nnn \l_Gtreg_list_seq {,} { #1 }
	\seq_use:Nn \l_Gtreg_list_seq {\hspace{-0.1em},\hspace{-0.18em} & }
}
\ExplSyntaxOff

\ExplSyntaxOn
\NewDocumentCommand{\GtregG}{m m m m}
{
	\Greg\!\left(\!\begin{smallmatrix}
		\Gtreg_print:n {#1} \\
		\Gtreg_print:n {#2}
	\end{smallmatrix}\,;\,\GtregG_print:n {#3}\,,\GtregG_print:n {#4}\,\Big|\,\SGroup\right)
}
\seq_new:N \l_GtregG_list_seq
\cs_new_protected:Npn \GtregG_print:n #1
{
	\seq_set_split:Nnn \l_GtregG_list_seq {,} { #1 }
	\seq_use:Nn \l_GtregG_list_seq {\hspace{-0.1em},\hspace{-0.18em} & }
}
\ExplSyntaxOff

\newcommand{\BW}{\mathrm{D}}

\usepackage[pdfencoding=auto,bookmarks=true,hyperfigures=true]{hyperref}%
\PassOptionsToPackage{unicode}{hyperref}
\providecommand{\hypersetup}[1]{}
\providecommand{\texorpdfstring}[2]{#1}
\hypersetup{plainpages=false}
\hypersetup{pdfpagemode=UseNone}
\hypersetup{bookmarksnumbered=true}
\hypersetup{pdfstartview=FitH}
\hypersetup{colorlinks=false}
\hypersetup{citebordercolor={0.7 0.7 1}}
\hypersetup{urlbordercolor={.4 .8 1}}
\hypersetup{linkbordercolor={1 .8 .6}}
\hypersetup{colorlinks=true, urlcolor=[rgb]{0.13,0.30,0.45}, linkcolor=[rgb]{0.13,0.30,0.45}, citecolor=[rgb]{0.55,0.0,0.05}}

\usepackage{tocloft}

\makeatletter
\newcommand{\appendixsection}[1]{%
  \refstepcounter{section}
  \section*{\thesection\quad#1}
  \addcontentsline{toc}{appendixsec}{\protect\numberline{\thesection}#1}
}
\newcommand{\appendixsubsection}[1]{
  \refstepcounter{subsection}
  \subsection*{\thesubsection\quad#1}
  \addcontentsline{toc}{appendixsubsec}{\protect\numberline{\thesubsection}#1}
}
\newcommand{\l@appendixsec}[2]{%
  \vspace{0.1ex}%
  \@dottedtocline{1}{\cftsubsecindent}{\cftsubsecnumwidth}{#1}{#2}%
}
\newcommand{\l@appendixsubsec}[2]{%
  \vspace{0.1ex}%
  \@dottedtocline{2}{\cftsubsubsecindent}{\cftsubsubsecnumwidth}{#1}{#2}%
}
\providecommand*{\toclevel@appendixsec}{1}
\providecommand*{\toclevel@appendixsubsec}{2}
\def\Hy@toc@appendixsec{section}
\def\Hy@toc@appendixsec{subsection}
\makeatother

\makeatletter
\let\@keywords\@empty
\let\@subject\@empty
\providecommand{\keywords}[1]{\gdef\@keywords{#1}}
\providecommand{\subject}[1]{\gdef\@subject{#1}}
\def\thetitle{\@title}
\def\theauthor{\@author}
\def\thesubject{\@subject}
\def\thedate{\@date}
\def\thekeywords{\@keywords}
\makeatother
\AtBeginDocument{
\hypersetup{pdftitle={\thetitle}}%
\hypersetup{pdfauthor={\theauthor}}%
\hypersetup{pdfsubject={\thesubject}}%
\hypersetup{pdfkeywords={\thekeywords}}%
}

\newif\ifnote\notetrue
\ifpublic
\notefalse
\fi

\newcommand{\ymnote}[1]{{\ifnote\textcolor{Maroon}{\normalfont\scriptsize\sffamily
\hspace{.1cm}YM: #1\hspace{.1cm}}\fi}} 

\let\Re\relax\DeclareMathOperator{\Re}{Re}
\let\Im\relax\DeclareMathOperator{\Im}{Im}

\let\Re\undefined\DeclareMathOperator{\Re}{Re}
\let\Im\undefined\DeclareMathOperator{\Im}{Im}

\DeclareMathOperator{\Ad}{Ad}
\DeclareMathOperator{\ad}{ad}

\DeclareMathOperator{\Li}{Li}
\DeclareMathOperator{\Lie}{Lie}

\DeclareMathOperator{\Gt}{\tilde{\Gamma}}

\DeclareMathOperator{\Sel}{S}
\DeclareMathOperator{\SelE}{S^E}

\newcommand{\SGroup}{\mathrm{G}}

\ProvideMathFix{rfrac,vfrac}
\newcommand{\iunit}{{\mathring{\imath}}}

\newcommand{\der}{\mathrm{d}}
\newcommand{\diff}[2][.]{\mathinner{\der#2\if #1.\else^{#1}\fi}}

\newcommand{\alg}[1]{\mathfrak{#1}}

\let\qed\relax\newcommand{\qed}
{\hfill\ensuremath{\Box}}

\newcommand{\pd}{\partial}

\newcommand{\vt}{\vartheta}

\newcommand{\eps}{\varepsilon}

\newcommand{\dd}{\mathrm{d}}

\newcommand{\zb}{\bar{z}}

\newcommand{\zC}{\mathbb C}

\newcommand{\zH}{\mathbb H}

\newcommand{\zP}{\mathbb P}

\newcommand{\zR}{\mathbb R}
\newcommand{\zZ}{\mathbb Z}

\newcommand{\cA}{\mathcal{A}}       
\newcommand{\cB}{\mathcal{B}}

\newcommand{\cE}{\mathcal{E}}

\newcommand{\cM}{\mathcal{M}}      
\newcommand{\cO}{\mathcal{O}}

\newcommand{\cU}{\mathcal{U}}

\newcommand{\cY}{\mathcal{Y}}

\newcommand{\sv}{\ensuremath{\text{sv}}}

\newcommand{\acyc}[0]{\mathfrak A}
\newcommand{\bcyc}[0]{\mathfrak B}

\newcommand{\Atxt}{$\mathfrak{A}$}
\newcommand{\Btxt}{$\mathfrak{B}$}

\newcommand{\gkern}[1]{g^{(#1)}}
\newcommand{\fkern}[1]{f^{(#1)}}

\renewcommand{\mid}{\,|\,}

\newcommand{\texteq}{\,{=}\,}

\ExplSyntaxOn
\NewDocumentCommand{\args}{o o}
 {
  	\IfValueTF{#1}
  	{
		\IfValueTF{#2}
		{
			\seq_set_from_clist:Nn \l_tmpa_seq { #1 }
  			\left( \seq_use:Nn \l_tmpa_seq { ,~ } ~;~ #2\right)
		}{
			\seq_set_from_clist:Nn \l_tmpa_seq { #1 }
  			\left( \seq_use:Nn \l_tmpa_seq { ,~ }\right)
		}
	}{
		\IfValueTF{#2}
		{
			\left( #2 \right)
		}{
			
		}
	}
 }
\ExplSyntaxOff

\newcommand{\szero}{e_0} 
\newcommand{\sone}{e_1} 

\newcommand{\drinfeld}[2]{\Phi(#1, #2)} 

\newcommand{\emplt}[2]{\Gamma(#1;#2|\tau)} 
\newcommand{\empltg}[2]{\tilde\Gamma(#1;#2|\tau)} 
\newcommand{\emplf}[4]{\Gamma\big(\begin{smallmatrix}#1\\#2\end{smallmatrix};#3\big|#4\big)}
\newcommand{\emplg}[4]{\tilde\Gamma\big(\begin{smallmatrix}#1\\#2\end{smallmatrix};#3\big|#4\big)}

\newcommand{\Aemzv}[1]{\omega_{\raisebox{-2.8pt}{\scriptsize$\acyc$}}(#1|\tau)} 
\newcommand{\Bemzv}[1]{\omega_{\raisebox{-2.8pt}{\scriptsize$\bcyc$}}(#1|\tau)} 
\newcommand{\coeffA}{\varpi_{\raisebox{-2.8pt}{\scriptsize$\acyc$}}} 
\newcommand{\coeffB}{\varpi_{\raisebox{-2.8pt}{\scriptsize$\bcyc$}}} 

\newcommand{\svcoeff}[2]{\mathbf{\Upsilon}(#1;#2|\tau)}
\newcommand{\emplcoeff}[2]{\Upsilon(#1;#2|\tau)}

\newcommand{\Drinfeld}{\Phi} 

\newcommand{\reverse}{\mathrm{R}} 

\newcommand{\KZB}{\omega_{\mathrm{KZB}}}

\newcommand{\alphatau}{\varphi_\tau}

\newcommand{\gzeroalph}{\cE}
\newcommand{\ellalph}{\cY}

\newcommand{\ZA}[1]{\mathtt{Z}_{\raisebox{-1.pt}{\scriptsize$\acyc$}}(#1|\tau)}
\newcommand{\ZB}[1]{\mathtt{Z}_{\raisebox{-1.pt}{\scriptsize$\bcyc$}}(#1|\tau)}

\newcommand{\diffform}{\phi}

\title{\textbf{\texorpdfstring{A construction of \\ single-valued elliptic polylogarithms}{A construction of single-valued elliptic polylogarithms}}}
\author{
Konstantin Baune, 
Johannes Broedel,
Yannis Moeckli
}
\date{\today}

\begin{document}

\pdfbookmark[1]{Title Page}{title} 
\thispagestyle{empty}
\vspace*{1.0cm}
\begin{center}%
  \begingroup\LARGE\bfseries\thetitle\par\endgroup
\vspace{1.0cm}

\begingroup\large\theauthor\par\endgroup
\vspace{9mm}
\begingroup\itshape
Institute for Theoretical Physics, ETH Zurich\\Wolfgang-Pauli-Str.~27, 8093 Zurich, Switzerland\\[4pt]
\par\endgroup
\vspace*{7mm}

\begingroup\ttfamily
baunek@ethz.ch, jbroedel@ethz.ch, moeckliy@ethz.ch 
\par\endgroup

\vspace*{2.0cm}

\textbf{Abstract}\vspace{5mm}

\begin{minipage}{13.4cm}
	We establish a general construction of single-valued elliptic polylogarithms as functions on the once-punctured elliptic curve. Our formalism is an extension of Brown's construction of genus-zero single-valued polylogarithms to the elliptic curve: the condition of trivial monodromy for solutions to the Knizhnik--Zamolodchikov--Bernard equation is expressed in terms of elliptic associators and involves two representations of a two-letter alphabet.\\
	Our elliptic single-valued condition reduces to Brown's genus-zero condition upon degeneration of the torus.
	We provide several examples for our construction, including the elliptic Bloch--Wigner dilogarithm.
\end{minipage} 
\end{center}
\vfill

\newpage
\setcounter{tocdepth}{2}
\tableofcontents


%
\section{Introduction}\label{sec:introduction}
Special functions are indispensable in the calculations of observables in physics. In particular, classical polylogarithms -- represented as iterated integrals over simple-pole differentials on the punctured Riemann sphere -- have proven very useful in numerous calculations, in particular for expressing scattering amplitudes in quantum field theory and string theory. Classical polylogarithms, however, are insufficient to encompass the entirety of observables: various Feynman diagrams and (worldsheet) correlators require polylogarithms defined on elliptic \cite{Levin,BrownLevin,Broedel:2017kkb,Broedel:2017siw,Broedel:2018qkq} and hyperelliptic curves \cite{Marzucca:2023gto,Duhr:2024uid} or Calabi--Yau manifolds \cite{Bonisch:2021yfw,Frellesvig:2023bbf,Frellesvig:2024rea,Klemm:2024wtd,Driesse:2024feo,Duhr:2025lbz,Duhr:2025xyy,e-collaboration:2025frv,Duhr:2025kkq,Duhr:2025ppd,Maggio:2025jel} (see \rcite{Bourjaily:2022bwx} for a review). In particular, loop amplitudes in string theory are formulated as correlation functions on Riemann surfaces of higher genus and are therefore naturally expressed in terms of higher-genus polylogarithms. 

The mathematical framework for polylogarithms on Riemann surfaces of higher genera has been extended substantially in the last decade. In general, polylogarithms can be constructed from a suitable connection form encoding the geometry of the surface in question and specifying appropriate boundary conditions. Solutions to the corresponding differential equation can be analytically continued by means of associators, which can be considered as arising from formally integrating the differential equation. 
Functional relations among those polylogarithms are then derived from exploring properties of the respective integration kernels implied by the connection form. 

For the elliptic case, the canonical connection form is the Knizhnik--Zamolodchikov--Bernard (KZB) connection, which is instrumental to the construction of polylogarithms on the torus \cite{Levin,BrownLevin}. The KZB connection is tied to the genus-one Fay relations \cite{Fay} and simultaneously to algebraic K-theory \cite{Bloch, GKZ, Broedel_2020}. Summarizing, using the associator formalism put forward in \rcites{CEE,Enriquez:EllAss} and its relation to braid groups, a versatile and complete framework for elliptic polylogarithms is available. Special values -- elliptic multiple zeta values (eMZVs) -- have been constructed in various contexts. While several results for the algebraic structure underlying eMZVs have been obtained \cite{Enriquez:Emzv,Matthes:Thesis,Broedel:2015hia,eMZVWebsite,katayama2025regularizationellipticmultiplezeta,lochak2020ellipticmultizetasellipticdouble}, the understanding of the space of eMZVs is not at the same level as for multiple zeta values (MZVs), which arise as special values of classical polylogarithms, see e.g.~\rcite{FresanGil}.

Stepping up to Riemann surfaces of higher genus, several connection forms -- and therefore different constructions of higher-genus polylogarithms -- are available \cite{EZ1,EZ2,EZ3,DHoker:2023vax,Ichikawa:2022qfx}. Functional relations for various kinds of higher-genus polylogarithms have been explored in \rcite{DHoker:2024ozn,Baune:2024ber,DHoker:2025dhv} and a generalization of multiple zeta values to surfaces of higher genera has been suggested in \rcite{Baune:2025sfy} based on the Schottky language established in \rcite{Baune:2024biq}.

\medskip

Polylogarithms of all genera can feature branch cuts, and elliptic and higher-genus polylogarithms are furthermore not necessarily periodic along the non-trivial cycles of the Riemann surface: accordingly, polylogarithms are multi-valued functions, and hence only well-defined on the universal covering space of the surface. Many physics observables, however, are single-valued objects. In order to properly represent those observables, the study of the space of single-valued polylogarithms is inevitable. 

Conveniently, single-valued polylogarithms can be expressed in terms of particular combinations of multi-valued polylogarithms and the associated multiple zeta values. Those combinations are designed in a way that monodromies around all non-trivial cycles of the Riemann surface cancel. The resulting set of single-valued functions obeys differential equations closely related to the KZB-type equation appropriate for the Riemann surface in question. Furthermore, it is desirable that the single-valued polylogarithms inherit a shuffle structure from their multivalued counterparts. 

The most prominent application of single-valued polylogarithms is in closed-string scattering: the relevant amplitudes can be expressed using the single-valued map \cite{Brown:2018omk,Brown:2019wna} or KLT relations \cite{Kawai:1985xq} from their open-string counterparts, which in turn are naturally expressed in terms of multi-valued polylogarithms. 

At genus zero, on the punctured Riemann sphere, the problem of constructing representations for single-valued polylogarithms was solved by Brown \cite{BrownSVMPL,Brownhyperlogs}. The condition of trivial monodromy of solutions to the holomorphic and anti-holomorphic version of the Knizhnik--Zamolodchikov equation has been expressed in terms of the Drinfeld associator using two different alphabets. This construction is as well the starting point for applying the single-valued map to open-string scattering amplitudes directly \cite{Brown:2019wna}. 

A general formalism for obtaining single-valued polylogarithms on Riemann surfaces of arbitrary genus is not known. In this article, we suggest a construction of single-valued elliptic polylogarithms, which is analogous to Brown's single-valued construction at genus zero: our formalism relies on expressing triviality of monodromies of solutions to the KZB equation in terms of elliptic associators, and connects two alphabets of algebraic letters related to the multi-valued solution and its complex conjugate. We point out the geometric origin of those relations and connect them via degeneration to the known genus-zero setup. In addition, we identify a modified KZB-type equation satisfied by our solution and prove that the resulting functions still obey the shuffle relations.

While only a few of single-valued elliptic polylogarithms have been constructed explicitly so far \cite{GKZ}, this is not true for the associated special values: for eMZVs or their representation in terms of iterated integrals of Eisenstein series, formulations of single-valued maps are known. Hereby iterated Eisenstein integrals of arbitrary depth are related to so-called modular graph functions, which are believed to span the space of single-valued elliptic polylogarithms and have been intensively investigated \cite{DHoker:2015wxz,Brown:mmv,Basu:2016kli,Panzer2019,DHoker:2019txf,Dorigoni:2019yoq,DHoker:2017pvk,Claasen:2025vcd,Gerken:2019cxz,DHoker:2020hlp,Dorigoni:2021jfr,Dorigoni:2021ngn}.

Extending the known construction of single-valued special values by a final integration with respect to one marked point $z$ on the torus leads to a construction of single-valued elliptic polylogarithms alternative to our construction: this formalism is going to be described in the article ``Elliptic modular graph forms, equivariant iterated integrals and single-valued elliptic polylogarithms'' \cite{SST} by Schlotterer, Sohnle and Tao appearing simultaneously to the current article.

\paragraph{Outline.} The article is structured as follows: in \secref{sec:genus0} we review Brown's construction of single-valued polylogarithms at genus zero. We then proceed in \secref{sec:genus1} by reviewing the KZB connection, elliptic associators and elliptic multiple polylogarithms. In \secref{sec:svempls} we describe our construction of single-valued elliptic polylogarithms by lifting Brown's genus-zero formalism to genus one using the elliptic associators to formulate a genus-one single-valued condition. In \secref{sec:examples} we provide several examples of single-valued elliptic polylogarithms, before concluding the article in \secref{sec:openqs} with a list of open questions.\\
Three appendices complement this article, where we note detailed expressions and derivations for several objects in our construction.  

\subsection*{Acknowledgments}
We are grateful to Manuel Berger, Egor Im, Stephan Stieberger and Federico Zerbini for various discussions, work on related projects and valuable feedback on a draft version of this manuscript. 
J.B.~would like to thank André Kaderli for conversations setting the stage for the problem solved within this article. 
The work of all authors
is partially supported by the Swiss National Science Foundation through the NCCR SwissMAP. 

\section{Single-valued polylogarithms at genus zero}\label{sec:genus0}
In this section we will review genus-zero polylogarithms and the construction of single-valued polylogarithms by Brown \cite{BrownSVMPL,Brownhyperlogs}, which we aim to lift to genus one in \secref{sec:svempls}. 

\subsection{Polylogarithms and MZVs at genus zero}\label{sec:mpls}
For the construction of \emph{multiple polylogarithms} (MPLs), also called \emph{hyperlogarithms} \cite{Brownhyperlogs,Goncharov:1998kja,Goncharov:2001iea}, we are dealing with the Knizhnik--Zamolodchikov-type equation
\begin{equation}\label{eqn:kzeq}
	\frac{\pd}{\pd z}G(z)=\left(\frac{\szero}{z}+\frac{\sone}{z-1}\right)G(z)
\end{equation}
on the punctured Riemann sphere $\zP^1(\zC)\setminus\{0,1,\infty\}$, where $\szero,\,\sone$ denote formal, non-commutative letters, which generate a free associative algebra\footnote{For the sake of brevity, we implicitly understand all algebras considered in this paper to be unital.} $\alg{f}_2(\gzeroalph)$ for $\gzeroalph\texteq\{\szero,\sone\}$.
Specifying the boundary condition $\lim_{z\to0}z^{-\szero}G(z)\texteq1$, the solution to this equation is unique \cite{Brownhyperlogs} and given by the formal series
\begin{equation}\label{eqn:mplgenseries}
	\begin{aligned}
		G(z)&=\sum_{w\in \gzeroalph^\times}G(w;z)\,w\\
		&=1+G(0;z)\,\szero+G(1;z)\,\sone+G(0,0;z)\,\szero^2+G(1,1;z)\,\sone^2\\
		&\quad+G(0,1;z)\,\szero\sone+G(1,0;z)\,\sone\szero+\ldots,
	\end{aligned}
\end{equation}
where $\gzeroalph^\times$ refers to the set of all words from letters $\szero,\,\sone$ (including the empty word) and the coefficients $G(w;z)$ are MPLs. They can be defined as iterated integrals by\footnote{We use the notation $G(e_{a_1},\ldots,e_{a_k};z)\,{\equiv}\, G(a_1,\ldots,a_k;z)$.}
\begin{align}\label{eqn:mpls}
	G(a_1,\ldots,a_k;z)=\int_0^z\omega_{a_1}(t)\,G(a_2,\ldots,a_k;t),
\end{align}
where $G(;z)\,{:=}\,1$, and 
\begin{equation}
	\omega_i(t)\texteq\frac{dt}{t-\sigma_i} \, ,\quad i\in\{0,1\}\, ,
\end{equation}
are simple-pole integration kernels corresponding to the punctures $\sigma_0\texteq0$ and $\sigma_1\texteq1$ of the Riemann sphere. Possible endpoint divergences of the integrals \eqref{eqn:mpls} are regularized by defining $G(0;z)\,{:=}\log z$ and using shuffle relations to transmit this prescription to all other divergent integrals \cite{Brown:2013qva}.

By means of the automorphism $z\,{\mapsto}\,1\,{-}\,z$ of the punctured Riemann sphere, we can find a solution $G_1(z)$ to the Knizhnik--Zamolodchikov equation such that $\lim_{z\rightarrow 1}(1-z)^{-e_1}G_1(z)\texteq1$. The ratio $G_1(z)^{-1}G(z)$ of the two solutions is independent of $z$ and called the \emph{Drinfeld associator}
\begin{equation}
	\drinfeld{e_0}{e_1}=\sum_{w\in\gzeroalph^\times}(-1)^{\#w}\zeta_w\,w,
\end{equation}
where $\zeta_w\texteq(-1)^{\#w}G(w;1)$ are (regularized) multiple zeta values (MZVs) obtained as special values of MPLs and $\#w$ is the number of letters $\sone$ within the word $w$.

It is possible to extend this formalism to both more marked points \cite{Brownhyperlogs} and more complex arguments \cite{BrownThesis}: in the case of a Riemann sphere with $N\,{+}\,1\,{\geq}\, 3$ punctures $\{z_0,\ldots,z_{N-1}\}\,{\cup}\,\{\infty\}$, the same mechanism applies when exchanging $\cE$ for an extended alphabet $\{e_0,\ldots,e_{N-1}\}$ assigning a letter to each puncture, and adjusting the KZ-equation accordingly to include all the marked points.
%

\subsection{Single-valued polylogarithms at genus zero}\label{sec:svmpls}
In this subsection, we review of the construction of single-valued MPLs (svMPLs) as established by Brown in \rcites{BrownSVMPL,Brownhyperlogs}. The MPLs reviewed in \secref{sec:mpls} are generalizations of the common logarithm obtained by iteratively integrating over simple poles: accordingly, they will exhibit branch cuts in the complex plane. Brown's formalism realizes single-valued versions of MPLs from sums and products of multi-valued MPLs (and MZVs) combined in a way that all branch cut contributions cancel.

Cancellation of all branch cut contributions is equivalent to demanding trivial monodromy of the generating function of single-valued polylogarithms along small paths around all possible branch points. Accordingly, the goal is to build combinations of holomorphic and anti-holomorphic (i.e.~complex conjugated) MPLs, so that all monodromies cancel. A possible representation of the generating series of svMPLs reads
\begin{align}\label{eqn:GG}
	\mathbf{G}(z)=G_{\gzeroalph}(z)\widetilde{\overline{{G}_{\gzeroalph'}(z)}},
\end{align}
where $\gzeroalph\texteq\{\szero,\sone\}$ and $\gzeroalph'\texteq\{\szero',\sone'\}$ are two different alphabets, and the operation $\ \widetilde{}\ $ reverses words. The subscripts $\gzeroalph$ and $\gzeroalph'$ denote the alphabet used in the expansion \eqref{eqn:mplgenseries}.

The monodromies $\cM_0$ around 0 and $\cM_1$ around 1 of the generating series of MPLs $G_{\gzeroalph}(z)$ read
\begin{subequations}
\begin{align}
	\cM_0G_{\gzeroalph}(z)&=G_{\gzeroalph}(z)e^{2\pi\iunit\szero},\\
	\cM_1G_{\gzeroalph}(z)&=G_{\gzeroalph}(z)\Drinfeld(\szero,\sone)^{-1}e^{-2\pi\iunit\sone}\Drinfeld(\szero,\sone).
\end{align}
\end{subequations}
In order for the monodromies to cancel in the combination $\mathbf{G}(z)\texteq G_{\gzeroalph}(z)\widetilde{\overline{G_{\gzeroalph'}(z)}}$, we thus have the following conditions between the alphabets $\gzeroalph$ and $\gzeroalph'$ (after linearizing the exponentials):
\begin{subequations}\label{eqn:svgenuszero}
	\begin{align}
		e_0'&=e_0,\\
		\Drinfeld(e_0,e_1)^{-1} e_1 \Drinfeld(e_0,e_1) &= \widetilde{\Drinfeld}(e_0',e_1')e_1'\widetilde{\Drinfeld}(e_0',e_1')^{-1},
	\end{align}
\end{subequations}
which we will refer to as the \emph{genus-zero single-valued condition}. Furthermore, we used that the Drinfeld associator has real coefficients, so that no complex conjugation is needed in the above formula.
The above relation can be solved iteratively to yield the unique solution
\begin{align}\label{eqn:e1p}
	e_1'&=e_1-2\zeta_3[e_0+e_1,[e_1,[e_0,e_1]]]+\ldots\,.
\end{align}
This change of alphabet has been explored using zeta generators in \rcites{Frost:2023stm,Frost:2025lre}. svMPLs $\mathbf{G}(w;z)$ are then obtained as coefficients of a word $w\,{\in}\,\gzeroalph^\times$ in the generating series \eqn{eqn:GG},
\begin{align}
	\mathbf{G}(z)=\sum_{w\in\gzeroalph^\times}\mathbf{G}(w;z)\,w.
\end{align}
where prior replacement of all primed letters with their unprimed versions via \eqn{eqn:e1p} is understood. The generating series $\mathbf{G}(z)$ still obeys the holomorphic Knizhnik--Zamolodchikov-type equation
\begin{align}
	\frac{\partial}{\partial z}\mathbf{G}(z)=\left(\frac{\szero}{z}+\frac{\sone}{z-1}\right)\mathbf{G}(z),
\end{align}
while the anti-holomorphic derivative reads
\begin{equation}
	\frac{\partial}{\partial \zb}\mathbf{G}(z)=\mathbf{G}(z)\left(\frac{\szero'}{\zb}+\frac{\sone'}{\zb-1}\right).
\end{equation}
The simplest example is the combination $\mathbf{G}(0;z)\texteq\log z\,{+}\,\log\zb\texteq\log|z|^2$, which is the single-valued version of the logarithm. The single-valued version of the dilogarithm $\mathrm{Li}_2(z)\texteq{-}\,G(0,1;z)$, known as the Bloch--Wigner dilogarithm \cite{Zagier:2007knq}, is given by $\mathrm{D}(z)\texteq\Im(\mathrm{Li}_2(z)\,{+}\,\log|z|\log(1{-}z))$, and can be identified (up to an additional single-valued term) as the single-valued version of $\mathrm{Li}_2(z)$ \cite{BrownSVMPL}:
\begin{equation}
\begin{aligned}
	\mathrm{Li}_2^{\sv}(z)&:=-\mathbf{G}(0,1;z)=2\iunit\Im(\mathrm{Li}_2(z)+\log|z|\log(1-z))-2\log|z|\log|1-z|\\
	&=2\iunit\,\mathrm{D}(z)-2\log|z|\log|1-z|.
\end{aligned}
\end{equation}

As pointed out before, MZVs are the special values of MPLs evaluated at $z\texteq1$. The construction of svMPLs now allows for the definition of single-valued MZVs (svMZVs) as the special values of svMPLs evaluated at $z\texteq1$
\begin{equation}
	\zeta^{\mathrm{sv}}_w:=(-1)^{\#w}\mathbf{G}(w;1),
\end{equation}
which were first studied in \rcite{Brown:2013gia}.

svMPLs and svMZVs make an appearance in many sophisticated physics examples, with the most prominent setup being tree-level closed-string amplitudes in flat space and AdS backgrounds, which have svMZVs as coefficients in their low-energy expansion \cite{Stieberger:2013wea,Stieberger:2014hba,Baune:2024uwj,Alday:2023mvu,Baune:2025hfu}. Furthermore, svMPLs show up in various quantum field theory amplitudes, see e.g.~\rcites{Dixon:2012yy,DelDuca:2016lad,DelDuca:2019tur}. Finally, the single-valued construction presented here can equivalently be phrased as a pairing of a differential form and a dual differential form as shown in \rcite{Brown:2018omk}, which was applied to tree-level string amplitudes in \rcite{Brown:2019wna}.

\section{KZB connection, elliptic polylogarithms and elliptic MZVs}\label{sec:genus1}

This section contains an overview of elliptic polylogarithms, elliptic associators and eMZVs. We will start by reviewing the KZB connection in \secref{sec:kzb} and its solution in \secref{sec:kzbsol}.
We are going to relate this solution to elliptic multiple polylogarithms in \secref{sec:empls} before defining and exploring elliptic associators in \secref{sec:ellass}.


\subsection{The KZB form}\label{sec:kzb}

Following Brown--Levin \cite{BrownLevin} 
and Matthes \cite{Matthes:Thesis}, 
we define the \emph{(elliptic) Knizhnik--Zamolodchikov--Bernard (KZB) form} as
\begin{equation}\label{eqn:BL}
	\KZB(z|\tau)= \Omega(z, \ad(b)|\tau)(a)-\nu\, b  \, ,
\end{equation}
for\footnote{In contrast to \rcite{BrownLevin}, we use the variant of the Jacobi theta function used in \rcite{Enriquez:Emzv} given by
	\begin{equation*}
		\theta(z|\tau)=\frac{e^{\pi\iunit z}-e^{-\pi\iunit z}}{2\pi\iunit}\prod_{n>0}\frac{\big(1-e^{2\pi\iunit(z+n\tau)}\big)\big(1-e^{2\pi\iunit(-z+n\tau)}\big)}{\big(1-e^{2\pi\iunit n\tau}\big)^2},
	\end{equation*}
	which is related to the odd Jacobi theta function via $\vartheta_1(z|\tau)\texteq2\pi\eta(\tau)^3\theta(z|\tau)$ for the Dedekind-$\eta$ function $\eta(\tau)$, so that the derivative of $\theta$ is normalized to $\theta'(0|\tau)\texteq1$.
	Also, in contrast to the conventions of \rcite{Matthes:Thesis}, we absorbed the additional pole-canceling factor $\alpha$ in the definition of $\Omega(z,\alpha|\tau)$.}
\begin{subequations}
	\begin{align}
		\Omega(z, \alpha|\tau) &= \exp(2\pi\iunit\,\alpha\, r)\,\alpha\, F(z,\alpha|\tau)\,\dd z = \exp\!\left(2\pi\iunit\,\alpha\frac{\Im(z)}{\Im(\tau)}\right)\alpha\frac{\theta(z+\alpha|\tau)}{\theta(z|\tau)\theta(\alpha|\tau)}\dd z,\\
		\nu &= 2\pi\iunit \,\dd r = \frac{2\pi\iunit}{\Im(\tau)} \dd\Im(z) \, ,\label{eqn:nu}
	\end{align}
\end{subequations}
where $z\texteq s \,{+}\, r\tau$ for $r,s\,{\in}\,[0,1]$ and $\tau\in\zH$, with $\zH$ being the Poincaré upper half plane, is the modular parameter determining the geometry of the elliptic curve. Moreover, $F(z,\alpha|\tau)$ denotes the Kronecker function. The KZB form is an integrable\footnote{A differential form $\diffform$ is integrable if $\dd\diffform\,{+}\, \diffform\wedge\diffform\texteq0$, where the second term is non-trivial if $\diffform$ takes values in a non-commutative algebra.} differential form on $E_\tau^\times\texteq\zC/(\zZ+\tau\zZ){\setminus}\{0\}$ valued in the associative algebra\footnote{Strictly speaking, $\omega_{\mathrm{KZB}}(z|\tau)$ is valued in the Lie-algebra $\Lie(a,b)$ freely generated by $a$ and $b$. However, we identify $\Lie(a,b)$ with its image in the universal enveloping algebra $\cU(\Lie(a,b))\,{\simeq}\, \alg{f}_2(\ellalph)$ via the canonical inclusion.} freely generated by $a$ and $b$ over $\zC$, which we denote by $\alg{f}_2(\ellalph)$ for $\ellalph\texteq\{a,b\}$.

We write the expansion of $\Omega(z, \ad(b)|\tau)(a)$ as 
\begin{equation}\label{eqn:fkerns}
	\Omega(z, \ad(b)|\tau)(a) = \sum_{k\geq 0}f^{(k)}(z|\tau)\,\dd z \ad^k(b)(a) \, ,
\end{equation}
where $f^{(k)}(z|\tau)$ is a function $E_\tau^\times\,{\rightarrow}\, \zC$ for all $k\,{\geq}\, 0$ and with $f^{(1)}(z|\tau)$ having a simple pole with residue $1$ at $z\texteq0$. The last property descends from the Kronecker function, which likewise exhibits a simple pole with unit residue at $z\texteq0$.

\paragraph{Universal KZB connection.}
The KZB form \eqref{eqn:BL} can be extended to the \emph{universal KZB form} by including terms proportional to $\dd\tau$ (see e.g.~\rcites{CEE,LevinRacinet}).
It is called \textit{universal}, as it describes the behavior of a solution under variations of both the point $z$ on the manifold and the geometric parameter $\tau$ of the surface. The additional term is conveniently represented as a sum over Eisenstein series and derivations of $\alg{f}_{2}(\ellalph)$.

In this article, we study solely variations of the variable $z$ on a torus of fixed geometry $\tau$, thus dealing with the KZB form as given in \eqn{eqn:BL}.

\subsection{The solution to the KZB equation}\label{sec:kzbsol}

Using the KZB form \eqref{eqn:BL}, one can formulate a generalization of the Knizhnik--Zamolodchikov equation on the punctured torus $E_\tau^\times$, see \rcites{BrownLevin,CEE,Matthes:Thesis}. In a second step, we describe a solution of this equation by a (suitably regularized) path-ordered exponential with values in the algebra $\alg{f}_2(\ellalph)$ \cite{Matthes:Thesis}. Expanding this solution in the generators of $\alg{f}_2(\ellalph)$ then gives rise to elliptic polylogarithms, which we will review in \secref{sec:empls}.

Let us start by describing the \emph{KZB equation} and its solutions, mainly following \rcites{Matthes:Thesis,Enriquez:Emzv}.
The KZB equation is the differential equation
\begin{equation}\label{eqn:KZBeq}
	\dd \Gamma(z|\tau) = -\KZB(z|\tau)\,\Gamma(z|\tau) \, ,
\end{equation}
where we consider the total differential to only include variations in $z$ and $\zb$, but not in $\tau$ (cf.~\eqn{eqn:BL}). Moreover, $\Gamma(z|\tau)$ denotes an $\alg{f}_2(\ellalph)$-valued function. The goal is to find an explicit solution to this equation, and in parallel to classify the space of possible solutions. In this regard, Matthes \cite{Matthes:Thesis} 
proved that there exists a unique solution $\Gamma(z|\tau)$ to the KZB equation \eqref{eqn:KZBeq} on $\{s+r\tau|s,r\,{\in}\,(0,1)\}\,{\subset}\,\zC$, such that 
\begin{equation}\label{eqn:asymp}
	\Gamma(z|\tau)\,{\sim}\, (-2\pi\iunit z)^{-[b,a]}\quad\text{as}\quad z\,{\rightarrow}\, 0.
\end{equation}
We will call $\Gamma(z|\tau)$ as defined above the \emph{elliptic KZB solution}. An explicit formula for the elliptic KZB solution $\Gamma(z|\tau)$ reads
\begin{equation}\label{eqn:formulaGenfunc}
	\Gamma(z|\tau) =\lim_{\eps\to0}\mathrm{Pexp}\!\left(-\int_\eps^z\KZB\right)(-2\pi\iunit\eps)^{-[b,a]} \, ,
\end{equation}
where the additional factor of $(-2\pi\iunit\eps)^{-[b,a]}$ for $\eps\in\zR_{>0}$ ensures the correct normalization by determining the regularization of possible end-point divergences. This particular regularization scheme is known as \emph{tangential basepoint regularization} (see e.g.~\rcite{EZ3} for a detailed account on regularization).
Moreover, any other solution $\Lambda(z|\tau)$ of the KZB equation can be obtained from $\Gamma(z|\tau)$ by multiplication with a $z$- and $\zb$-independent factor 
\begin{equation}\label{eqn:KZB2ndsol}
	\Lambda(z|\tau) = \Gamma(z|\tau)\,S(\tau).
\end{equation}
This can be seen, by calculating that $\dd (\Gamma(z|\tau)^{-1}\Lambda(z|\tau))\texteq0$, using that both $\Gamma(z|\tau)$ and $\Lambda(z|\tau)$ are solutions of the KZB equation (and that the differential does not act on $\tau$). Thus, we find that the combination $\Gamma(z|\tau)^{-1}\Lambda(z|\tau)$ must be equal to a function independent of $z$ and $\zb$, from which \eqn{eqn:KZB2ndsol} follows. 

\paragraph{Shuffle relations and a Hopf algebra.} \label{paragraph:Hopf}

It can be shown that the coefficients of this solution satisfy the typical \emph{shuffle relations}, which is to be expected since the coefficients are expressed as (suitably regularized) homotopy invariant combinations of iterated integrals. However, let us elaborate a bit more on this fact as it introduces a convenient algebraic language for dealing with polylogarithms in general.

Recall that $\Gamma(z|\tau)$ is valued in the algebra $\alg{f}_2(\ellalph)$, which has a rich algebraic structure. In particular, $\alg{f}_2(\ellalph)$ has a natural Hopf algebra structure w.r.t.~concatenation product, for which the coproduct $\Delta$ is given by requiring the generators $a$ and $b$ to be primitive, i.e.
\begin{equation}\label{eqn:coproduct}
	\Delta(a) = a\otimes1 + 1\otimes a\, , \quad \Delta(b) = b\otimes1 + 1\otimes b,
\end{equation} 
and the antipode $\varsigma$ is given by the unique anti-homomorphism induced by
\begin{equation}\label{eqn:antipode}
\varsigma(a)=-a\, , \quad \varsigma(b)=-b \, .
\end{equation} 
For further details, consider e.g.~\rcite{Chari:1994pz}. It can now be shown that the elliptic KZB solution is group-like w.r.t.~this coproduct, i.e.~$\Delta(\Gamma(z|\tau))\texteq\Gamma(z|\tau)\otimes\Gamma(z|\tau)$ (see e.g.~\rcite[Prop.~A.2.3]{Matthes:Thesis}). A result of Brown \cite[Cor.~3.5]{Brownhyperlogs} then in turn implies directly that its coefficients satisfy the shuffle relations.

\paragraph{Homotopy invariant iterated integrals.}

Since the elliptic KZB solution is valued in the algebra $\alg{f}_2(\ellalph)$, it can be expanded in the generators $a$ and $b$. To start with, notice that \eqn{eqn:formulaGenfunc} implies that we can write
\begin{equation}\label{eqn:xexpansion}
	\begin{aligned}
		\Gamma(z|\tau)=1&+\int_0^z\nu(t|\tau) \, b - \int_0^zf^{(0)}(t|\tau)\,\dd t \, a
		- \int_0^zf^{(1)}(t|\tau)\,\dd t\,\ad(b)(a) \\
		&+\int_0^zf^{(0)}(t_1|\tau)\dd t_1\int_0^{t_1}f^{(0)}(t_2|\tau)\,\dd t_2\, a^2 + \int_0^z \nu(t_1|\tau)\int_0^{t_1}\nu(t_2|\tau)\, b^2 \\
		&-\int_0^z\nu(t_1|\tau)\int_0^{t_1}f^{(0)}(t_2|\tau)\,\dd t_2\, ba - \int_0^z f^{(0)}(t_1|\tau)\,\dd t_1\int_0^{t_1}\nu(t_2|\tau)\, ab \\
		&+ \ldots
	\end{aligned}
\end{equation}
upon expansion of the KZB form (as given in \eqn{eqn:BL} ff.), where the integrals above are understood to be tangentially basepoint regularized as described above. Finally, upon further expanding the (nested) commutators $\ad(b)^k(a)$, $k\geq0$, appearing in the KZB form, we finally obtain
\begin{equation}\label{eqn:hicoeff}
	\begin{aligned}
		\Gamma(z|\tau)&=1+\int_0^z\nu(t|\tau)\, b - \int_0^zf^{(0)}(t|\tau)\,\dd t\, a \\
		&\quad\quad\!+\int_0^zf^{(0)}(t_1|\tau)\dd t_1\int_0^{t_1}f^{(0)}(t_2|\tau)\,\dd t_2\, a^2 + \int_0^z \nu(t_1|\tau)\int_0^{t_1}\nu(t_2|\tau)\, b^2 \\
		&\quad\quad\!+ \left(-\int_0^zf^{(1)}(t|\tau)\,\dd t-\int_0^z\nu(t_1|\tau)\int_0^{t_1}f^{(0)}(t_2|\tau)\,\dd t_2\right) ba \\
		&\quad\quad\!+ \left(\int_0^zf^{(1)}(t|\tau)\,\dd t-\int_0^z f^{(0)}(t_1|\tau)\,\dd t_1\int_0^{t_2}\nu(t_2|\tau)\right) ab + \ldots \\
		&=:\sum_{w\in \ellalph^\times}\emplcoeff{w}{z}\, w \, .
	\end{aligned}
\end{equation}
Since the KZB form is integrable, the coefficients $\emplcoeff{w}{z}$ of the elliptic KZB solution define homotopy invariant (combinations of) iterated integrals \cite{BrownLevin}. Despite the nice properties of the coefficients $\emplcoeff{w}{z}$ such as shuffle relations and homotopy invariance, these functions are not naturally defined on the torus as they generically feature branch cuts and non-trivial monodromies around non-contractible cycles of the torus. The goal of this article is to remedy this flaw by introducing a suitable modification of these functions, that is naturally defined on $E_\tau^\times$, i.e.~does feature neither branch cuts nor non-trivial monodromies around non-contractible cycles.

\subsection{Elliptic multiple polylogarithms}\label{sec:empls}

In this subsection, we want to draw the connection to the notion of \emph{elliptic multiple polylogarithms (eMPLs)} typically appearing in the literature (see e.g.~\rcites{Broedel:2017kkb,Matthes:Thesis,EZ3,Broedel:2019gba}). Typically, eMPLs are defined as iterated integrals
\begin{equation}\label{eqn:eMPLs}
	\emplg{n_1 & \cdots & n_r}{a_1 & \cdots & a_r}{z}{\tau}=\int_0^z\gkern{n_1}(t-a_1|\tau)\,\emplg{n_2 & \cdots & n_r}{a_2 & \cdots & a_r}{t}{\tau}\,\dd t\, ,
\end{equation}
where the functions $\gkern{n}(z|\tau)$, $n\geq0$, can be recursively defined by the equalities \cite{Broedel:2014vla}
\begin{equation}\label{eqn:fkern}
	\fkern{n}(z|\tau)=\sum_{k=0}^{n}\frac{1}{k!}\left(\frac{2\pi\iunit\Im(z)}{\Im(\tau)}\right)^k\gkern{n-k}(z|\tau).
\end{equation}
These functions are meromorphic and quasi-periodic, i.e.~only defined on a universal covering space of $E_\tau^\times$. Meromorphicity directly implies homotopy invariance of the iterated integrals defined in \eqn{eqn:eMPLs}. However, the fact that the functions $\gkern{n}(z|\tau)$ are only defined on a covering space introduces several complications when attempting to construct single-valued eMPLs, which is why we will mainly deal with the functions $\emplcoeff{w}{z}$ defined in \eqn{eqn:hicoeff}.

Using identity \eqref{eqn:fkern}, we can relate the iterated integrals \eqref{eqn:eMPLs} to the functions\footnote{These integrals can be obtained from expanding the path-ordered exponential using the (non-integrable) connection form $\Omega(z,\ad(b)|\tau)(a)$ (cf.~\eqn{eqn:fkerns}) with respect to the letters $x_k\texteq\ad(b)^k(a)$, see e.g.~\rcite{Broedel:2014vla}.}
\begin{equation}\label{eqn:ellfint}
	\emplf{n_1 & \cdots & n_r}{a_1 & \cdots & a_r}{z}{\tau}=\int_0^z\fkern{n_1}(t-a_1|\tau)\,\emplf{n_2 & \cdots & n_r}{a_2 & \cdots & a_r}{t}{\tau}\,\dd t,
\end{equation}
which we call \emph{elliptic $f$-integrals}, and it can in fact be shown that the eMPLs \eqref{eqn:eMPLs} span the same function space as the elliptic $f$-integrals \eqref{eqn:ellfint} upon adding polynomials in $\zb$ \cite{EZ3}. In the following, we will exclusively work with $a_1\texteq\ldots\texteq a_r\texteq0$ and thus omit these variables, using the notation $\emplf{n_1&\cdots&n_r}{0&\cdots&0}{z}{\tau}\,{\equiv}\,\emplt{n_1,\ldots,n_r}{z}$ (and likewise for $\tilde\Gamma$). This is because we are only dealing with only one puncture of the torus, which we chose to be at $z\texteq0$.

The crucial drawback of the elliptic $f$-integrals is that these are in general not homotopy invariant since the functions $f^{(k)}(z|\tau)$ are not holomorphic (and thus the forms $f^{(k)}(z|\tau)\dd z$ not necessarily closed). To illustrate this, consider the integral of $f^{(1)}(z|\tau)$ along a path $\gamma$. Let moreover $\delta$ be another path in the same (endpoint) homotopy class as $\gamma$. Then by Stokes' theorem
\begin{equation}
	\int_{\gamma}f^{(1)}(z|\tau)\,\dd z - \int_{\delta}f^{(1)}(z|\tau)\,\dd z = \int_{\mathrm{Int}(\gamma-\delta)}\dd (f^{(1)}(z|\tau)\,\dd z) = -\frac{\pi}{\Im(\tau)}\int_{\mathrm{Int}(\gamma-\delta)}\,\dd \zb\wedge \dd z \, ,
\end{equation}
where $\mathrm{Int}(\gamma\,{-}\,\delta)$ refers to the interior of the region bounded by the path $\gamma\,{-}\,\delta$. This generally does not vanish, which implies that the integral is not homotopy invariant. Homotopy invariance is recovered upon including the differential form $\nu$, which returns the coefficients $\emplcoeff{w}{z}$ obtained by expanding the elliptic KZB solution. 

However, as a final remark, notice that it is in general not possible to lift an elliptic $f$-integral to a coefficient $\emplcoeff{w}{z}$ for some $w\,{\in}\,\ellalph^\times$ in a unique way. This can be seen e.g.~in \eqn{eqn:hicoeff}, where it is evident that the elliptic $f$-integral $\emplt{1}{z}$ can be lifted to either $\emplcoeff{ba}{z}$ or $\emplcoeff{ab}{z}$.

If we can relate a given eMPL $\emplg{n_1 & \cdots & n_r}{a_1 & \cdots & a_r}{z}{\tau}$ or elliptic $f$-integral $\emplf{n_1 & \cdots & n_r}{a_1 & \cdots & a_r}{z}{\tau}$ (including the choice of an explicit path due to the lack of homotopy invariance in this case) to the coefficients $\emplcoeff{w}{z}$, our construction of single-valued versions of $\emplcoeff{w}{z}$ thus immediately yields single-valued versions of eMPLs and elliptic $f$-integrals. We discuss some examples in \secref{sec:examples} below.

\paragraph{Elliptic multiple zeta values.}\label{sec:emzvs}
In analogy to MZVs being special values of MPLs at $z\texteq1$, we can define \emph{elliptic multiple zeta values} (eMZVs) as special values of elliptic $f$-integrals, where the integration path is taken to be the \Atxt- or \Btxt-cycle\footnote{Note the reversed order of the arguments between eMPLs and eMZVs, which is the usual convention in the literature.} \cite{Matthes:Thesis,Enriquez:Emzv}:
\begin{subequations}\label{eqn:emzvs}
	\begin{align}
		\Aemzv{n_1,\ldots,n_r}&=\emplt{n_r,\ldots, n_1}{1},\\
		\Bemzv{n_1,\ldots,n_r}&=\emplt{n_r,\ldots, n_1}{\tau}.
	\end{align}
\end{subequations}
As a small remark, notice that the eMZVs w.r.t.~the $\acyc$-cycle can equivalently be defined via the eMPLs by replacing $\Gamma$ with $\tilde\Gamma$ as the $\acyc$-cycle can be taken as the real interval $[0,1]$ on which the functions $\fkern{k}(z|\tau)$ and $\gkern{k}(z|\tau)$ coincide. This way, we can see that they are actually given as special values of homotopy invariant iterated integrals. This is different for the $\bcyc$-cycle eMZVs, where they depend on the specific choice of path between $0$ and $\tau$ due to the lack of homotopy invariance of the elliptic $f$-integrals.

These special values are important when defining associators on the genus-one surface and later when formulating a single-valued condition for eMPLs. Their relations and $q$-expansions have been studied in \rcites{Broedel_2020,Matthes:Thesis,Broedel:2015hia,Matthes:Decomposition,Broedel:2018izr} and they have been applied in the physics context for example in one-loop open-string amplitudes in \rcites{Broedel:2019gba,Broedel:2014vla,Broedel:2018izr,Mafra:2019ddf,Mafra:2019xms,Broedel:2017jdo,Broedel:2020tmd}.

\subsection{Elliptic associators} \label{sec:ellass}
In \secref{sec:svmpls}, the genus-zero monodromy conditions were formulated in terms of the Drinfeld associator. In turn, the associator (completely) describes the behavior of the generating function of MPLs \eqref{eqn:mplgenseries} when analytically continued around a non-contractible cycle in the system $\zP^1(\zC){\setminus}\{0,1,\infty\}$. Therefore, in order to describe a generalization of the single-valued procedure, we first require a generalization of the concept of associators to the (punctured) torus $E_\tau^\times$. 
\begin{figure}
	\centering
	\mpostuse{ellipticassociators}
	\caption{Graphical representation of the A- and B-cycle associators defined by Enriquez \cite{Enriquez:EllAss}. The extra phase for the A-associator reflects our convention chosen in this article: the elliptic associators always interpolate between the same tangential basepoints. For better visibility, even though both associators are defined with respect to the same tangential basepoint, the tangential basepoint for the B-cycle associator is illustrated as pointing horizontally to the right, while the tangential basepoint of the A-cycle associator is drawn as pointing vertically down on the torus in the second picture. The half-circles in this case indicate that the associators interpolate between the same tangential basepoint.}
	\label{fig:ellipticassociators}
\end{figure}

In order to achieve such a generalization, recall that the fundamental group of the torus can be described as freely generated by the homotopy classes of the two non-contractible cycles $\acyc$ and $\bcyc$ (cf.~\figref{fig:ellipticassociators}). Accordingly, following \rcites{Enriquez:EllAss,Enriquez:Emzv}, we define the \emph{elliptic associators} $A(\tau)$ and $B(\tau)$ to be the group-like series in $\alg{f}_2(\ellalph)$ defined by\footnote{Notice that, due to $\Gamma(z|\tau)$ being a solution to the KZB equation~\eqref{eqn:KZBeq}, the combinations used in \eqn{eqn:ellass} are indeed independent of $z$ according to \eqn{eqn:KZB2ndsol}. The fact that these series are group-like is immediate from the compatibility axioms of a Hopf algebra and using that $\Gamma(z|\tau)$ is group-like as well.}
\begin{subequations}\label{eqn:ellass}
	\begin{align}
		A(\tau) &= \Gamma(z|\tau)^{-1}\Gamma(z+\acyc|\tau) = \Gamma(z|\tau)^{-1}\Gamma(z+1|\tau) \, , \\
		B(\tau) &= \Gamma(z|\tau)^{-1}\Gamma(z+\bcyc|\tau) = \Gamma(z|\tau)^{-1}\Gamma(z+\tau|\tau) \, ,
	\end{align}
\end{subequations}
where $\Gamma(z|\tau)$ is the elliptic KZB solution. We will refer to $A(\tau)$ as the \emph{A-associator}, whereas $B(\tau)$ will be referred to as the \emph{B-associator}. Using \eqn{eqn:formulaGenfunc}, we can express the elliptic associators as
\begin{subequations}
	\begin{align}
		\label{eqn:formulaAass}A(\tau) &= \lim_{\eps\rightarrow 0}(-2\pi\iunit\eps)^{[b,a]}\,\mathrm{Pexp}\!\left(\int_\eps^z\KZB\right)\mathrm{Pexp}\!\left(-\int_\eps^{z+1}\KZB\right)(-2\pi\iunit\eps)^{-[b,a]} \notag\\
		&= \lim_{\eps\rightarrow 0}(-2\pi\iunit\eps)^{[b,a]}\,\mathrm{Pexp}\!\left(\int_\eps^z\KZB\right)\mathrm{Pexp}\!\left(-\int_{\eps-1}^z\KZB\right)(-2\pi\iunit\eps)^{-[b,a]}\notag \\
		&= \lim_{\eps\rightarrow 0}(-2\pi\iunit\eps)^{[b,a]}\,\mathrm{Pexp}\!\left(-\int_\eps^{1+\eps}\KZB\right)(-2\pi\iunit\eps)^{-[b,a]} \, , \\
		\label{eqn:formulaBass}B(\tau) &= \lim_{\eps\rightarrow 0}(-2\pi\iunit\eps)^{[b,a]}\,\mathrm{Pexp}\!\left(-\int_\eps^{\tau+\eps}\KZB\right)(-2\pi\iunit\eps)^{-[b,a]} \, .
	\end{align}
\end{subequations}
Notice that the direction of $\eps$ in these formulas, which has been fixed to be $\eps\in\zR_{>0}$ in \secref{sec:kzbsol}, again determines the tangential basepoint used in the regularization scheme. Our choice follows the original definition of elliptic associators by Enriquez \cite{Enriquez:EllAss} and can be interpreted as defining the A- and B-associator as the parallel transport of the elliptic KZB solution between the tangential basepoints $0$ and $1$ and $0$ and $\tau$ respectively, where the tangent vector is chosen to point along the positive real direction in all cases. This convention is also illustrated in \figref{fig:ellipticassociators} as the half-circle at the endpoint $1$ reflecting the necessity to arrive at the same tangent vector after parallel transporting.

As for the elliptic KZB solution, we can formally expand the elliptic associators in the generators of the algebra $\alg{f}_2(\ellalph)$. We can do this either in the alphabet $\ellalph$, resulting in words of $a$ and $b$, or choose to expand in (nested) commutators\footnote{Whereas this is an equivalent choice for $A(\tau)$, the appearance of the isolated letters $b\texteq x_{-1}$ in $B(\tau)$ renders this choice of alphabet inequivalent to $\ellalph$. This can be seen immediately when noticing that $b\texteq x_{-1}$ introduces relations in this alphabet as e.g.~$x_1\texteq x_{-1}x_0\,{-}\,x_0x_{-1}$, while $\ellalph$ constitutes an independent set of words. We will comment on this subtlety in more detail below.} $x_k\,{:=}\,\ad(b)^k(a)$ for $k\geq-1$, where we conveniently define $x_{-1}\texteq\ad(b)^{-1}(a)\,{:=}\,b$. This results in two different sets of expansion coefficients, i.e.
\begin{subequations}
\begin{align}
	\label{eqn:expAass}A(\tau) &=\sum_{w\in\ellalph^\times}\ZA{w}\,w \notag\\
	&= \sum_{n\geq 0} \sum_{k_1,\ldots,k_n\geq 0} \coeffA(k_1,\ldots,k_n|\tau)x_{k_1}\cdots x_{k_n},\\
	\label{eqn:expBass}B(\tau) &=\sum_{w\in\ellalph^\times}\ZB{w}\,w \notag\\
	&= \sum_{n\geq 0}\sum_{k_1,\ldots,k_n\geq -1} \coeffB(k_1,\ldots,k_n|\tau)x_{k_1}\cdots x_{k_n}. 
\end{align}
\end{subequations}
Note that according to \eqn{eqn:formulaAass} and \eqn{eqn:formulaBass}, the coefficients $\ZA{w}$ and $\ZB{w}$ are given by homotopy invariant (combinations of) iterated integrals, hence we can choose the $\acyc$- and $\bcyc$-cycles to coincide with the real interval $[0,1]$ and the straight line connecting $0$ and $\tau$, respectively (cf.~\figref{fig:ellipticassociators}). This configuration of cycles implies vanishing of the differential form $\nu$ on the $\acyc$-cycle (cf.~\eqn{eqn:nu}), but not necessarily on the $\bcyc$-cycle.

This implies a crucial difference between the A- and B-associator: since $\nu$ vanishes on the $\acyc$-cycle, it can be seen that the coefficients $\ZA{w}$ are proportional to $\coeffA(k_1, \ldots,k_n|\tau)$ (with proportionality factors $\pm1$ from expanding the commutators) for some $k_1,\ldots,k_n\,{\geq}\,0$ such that the word $w$ appears in $x_{k_1}\cdots x_{k_n}$ upon expanding the commutators. This matching is unique in the sense that every word $w\,{\in}\,\ellalph^\times$ corresponds to exactly one (nested) commutator, i.e.~word in $x_k$ for $k\,{\geq}\,0$, hence we can relate the coefficients directly. For the B-associator, this correspondence is broken due to the non-vanishing of $\nu$ on the $\bcyc$-cycle. This implies non-trivial relations between $\ZB{w}$ and $\coeffB(k_1,\ldots,k_n|\tau)$, $k_1,\ldots,k_n\,{\geq}\,0$. For example, at length two, we find
\begin{equation}
	\begin{aligned}\label{eqn:coeffBrel}
		\ZB{a^2} &= \coeffB(0,0|\tau) \, ,\\
		\ZB{ab} &= - \coeffB(1|\tau) + \coeffB(0,-1|\tau) \, ,\\
		\ZB{ba} &= \coeffB(1|\tau) + \coeffB(-1,0|\tau) \, ,\\
		\ZB{b^2} &= \coeffB(-1,-1|\tau) \, .
	\end{aligned}
\end{equation}
Clearly, the set of equations \eqref{eqn:coeffBrel} cannot be resolved to express the coefficients $\coeffB(k_1,\ldots,k_n|\tau)$ in terms of the coefficients $\ZB{w}$ entirely. Note that this is in complete analogy with \eqn{eqn:hicoeff} expressing the homotopy invariant combinations in terms of iterated integrals over the set of differential forms $\{\nu,f^{(k)}(z|\tau)\,\dd z|k\,{\geq}\,0\}$, whereas the single iterated integrals themselves are not homotopy invariant. There, we have seen that the elliptic $f$-integrals cannot be uniquely lifted to a homotopy invariant combination $\emplcoeff{w}{z}$ (cf.~\secref{sec:empls}). This is the exact same situation here, where the coefficients $\coeffB(k_1,\ldots,k_n|\tau)$, $k_1,\ldots,k_n\,{\geq}\,{-}\,1$, essentially correspond to special values of the elliptic $f$-integrals whereas the coefficients $\ZB{w}$ correspond to special values of the $\emplcoeff{w}{z}$. Despite the coefficients $\ZB{w}$ being generally better behaved, we will express many results in this work in terms of the $\coeffB(k_1,\ldots,k_n|\tau)$ since they are closer to the notion of elliptic B-MZVs in the literature \cite{Enriquez:Emzv}. We will elaborate on this relation in a bit more detail below.

Finally, we can make use of the formulas \eqn{eqn:formulaAass} and \eqn{eqn:formulaBass} to calculate the associators explicitly. The result to second order reads
\begin{subequations}\label{eqn:assexpansion}
\begin{align}
	A(\tau) &= 1-a+\pi\iunit[b,a]+\frac12 a^2 + \ldots \, , \\
	B(\tau) &= 1+2\pi\iunit b-\tau a-\pi\iunit\,[b,a] - \pi\iunit\tau\{b,a\}+\frac{(2\pi\iunit)^2}{2}b^2+\frac{\tau^2}{2} a^2 + \ldots \, ,
\end{align}
\end{subequations}
where we denoted the anti-commutator by $\{b,a\}$. These formulas again illustrate the caveats described above. It can clearly be seen that the A-associator is purely expressed in terms of commutators whereas this property is lost in the B-associator due to the appearance of the anti-commutator.

\paragraph{Modular $S$-relation.} Recall that the action of the modular group $\mathrm{SL}_2(\zZ)$ maps a given torus to an equivalent torus with a potentially different basis of the (first) homology group $H_1(\zZ,E_\tau^\times)$, see e.g.~\rcite{Bobenko:2011}. In particular, the modular $S$-transformation defined via $S(\tau)\texteq-\tau^{-1}$ can be interpreted as swapping the (homology classes of) the non-contractible cycles $\acyc$ and $\bcyc$. This symmetry descends to the elliptic associators in the sense that they satisfy the \emph{modular S-relation} \cite{Enriquez:Emzv}: define the map $\alphatau:\alg{f}_2(\ellalph)\rightarrow\alg{f}_2(\ellalph)$ through $\alphatau(b)\texteq{-}\,\tau b,$ $\alphatau(a)\texteq2\pi\iunit b\,{-}\,\tau^{-1} a$. Then it holds that
\begin{equation}\label{eqn:relationAB}
\begin{aligned}
	A\!\left(-\tfrac1\tau\right)&=\Ad\!\left(\!\left(-\tfrac1\tau\right)^{[b,a]}\right)\circ\alphatau(B(\tau)^{-1})\\
	&=\left(-\tfrac1\tau\right)^{[b,a]}\alphatau(B(\tau)^{-1})\left(-\tfrac1\tau\right)^{-[b,a]},
\end{aligned}
\end{equation}
where $\Ad$ is the adjoint action defined as shown in second line of the above equation. This relation allows for (uniquely) expressing the coefficients $\ZB{w}$ in terms of a combination of coefficients $\mathtt{Z}_{\raisebox{-1.pt}{\scriptsize$\acyc$}}(w|{-}\tau^{-1})$, which consequently also relates the coefficients $\coeffA\left(k_1,\ldots,k_n|{-}\tau^{-1}\right)$ and $\coeffB(k_1,\ldots,k_n|\tau)$. However, due to the caveats mentioned above, the relations between the latter cannot uniquely be solved for the coefficients $\coeffB(k_1,\ldots,k_n|\tau)$ solely in terms of the coefficients $\coeffA(k_1,\ldots,k_n|-\tau^{-1})$. Similar notions of expansion coefficients were discussed in ref.~\cite[Sec.~3]{Matthes:Thesis}.

Using the expansions \eqn{eqn:assexpansion}, we can now verify the modular relation \eqn{eqn:relationAB} up to the order computed above:
\begin{align}
	&\Ad\left((-\tau^{-1})^{[b,a]}\right)\circ\alphatau\left(B(\tau)^{-1}\right)\notag\\
	=&\Ad\left((-\tau^{-1})^{[b,a]}\right)\circ\alphatau\bigg(1-2\pi\iunit b+\tau a+\pi\iunit\,[b,a]-2\pi\iunit\frac{\tau}{2}\{b,a\}+\frac{(2\pi\iunit)^2}{2}b^2+\frac{\tau^2}{2} a^2 + \ldots\bigg)\notag \\
	=&\Ad\left((-\tau^{-1})^{[b,a]}\right)\circ\left(1-a+\pi\iunit\,[b,a]+\frac12 a^2 + \ldots\right) \notag\\
	=&\,A(-\tau^{-1}) \, ,
\end{align}
where in the last step we have used that the adjoint action does not contribute up to second order. Going to higher length, one can extract further relations between these coefficients, which are (partially) collected in \appref{app:coeffs}.

\paragraph{Relation to eMZVs.} Finally, we want to draw the connection of the coefficients appearing in the expansions \eqn{eqn:expAass} and \eqn{eqn:expBass} of the elliptic associators to the eMZVs defined in \eqn{eqn:emzvs} and commonly referred to in the literature \cite{Matthes:Thesis,Enriquez:Emzv}. Enriquez' A-associator discussed above can be expressed in terms of \Atxt-cycle eMZVs via \cite{Matthes:Thesis}
\begin{equation}
	\begin{aligned}\label{eqn:Aexpansion}
		A(\tau) &= e^{\pi\iunit \ad(b)(a)} \sum_{n\geq 0} (-1)^n\sum_{k_1,\ldots,k_n\geq 0} \Aemzv{k_n,\ldots,k_1}\ad(b)^{k_1}(a)\cdots\ad(b)^{k_n}(a) \, .
	\end{aligned}
\end{equation}
Upon expanding the above relation, we can find relations among the coefficients $\coeffA(k_1,\ldots,k_n|\tau)$ and the eMZVs, several of them are collected in \eqn{eqn:exampleseMZVs} below.

The B-associator \eqref{eqn:expBass} is however not exactly a generating series of the \Btxt-cycle eMZVs according to \rcite{Enriquez:Emzv}, but its coefficients can be deduced in terms of \Atxt-cycle eMZVs (evaluated at ${-}\tau^{-1}$) using \eqn{eqn:relationAB}.

Examples of low weights include:
\begin{align}\label{eqn:exampleseMZVs}
	&\begin{aligned}
		\coeffA(0,1|\tau)&=\Aemzv{1,0}, &&\qquad\qquad\coeffA(1,0|\tau)=\Aemzv{0,1}-\pi\iunit\,\Aemzv{0},\\
		\coeffA(1,1|\tau)&=\Aemzv{1,1}-\pi\iunit\,\Aemzv{1}-\tfrac{\pi^2}{2},
		&&\qquad\qquad\hspace{0.45ex}\coeffB\hspace{-0.25ex}(-1|\tau)=-2\pi\iunit\coeffA(0|{-}\tau^{-1}), 
	\end{aligned}\notag\\
	&\coeffB\hspace{-0.25ex}(0,1|\tau)=\tau\coeffA(1|{-}\tau^{-1})+\tau\coeffA(0,1|{-}\tau^{-1})-\tau\log({-}\tau^{-1}).
\end{align}

\section{A construction of single-valued polylogarithms at genus one}\label{sec:svempls}
In this section, we are going to present our construction of \emph{single-valued elliptic polylogarithms} (sveMPLs), which lifts Brown's single-valued construction (reviewed in \secref{sec:svmpls}) from genus zero to genus one.

To formulate an elliptic single-valued condition analogous to the genus-zero condition \eqref{eqn:svgenuszero}, we will use elliptic associators (as reviewed in \secref{sec:ellass}) to relate two different alphabets of algebraic letters, which will then ensure trivial monodromies of our sveMPLs. In \secref{sec:degen}, we will show that the single-valued condition along the A-cycle yields the known genus-zero single-valued condition if the elliptic curve is degenerated to the torus. 

\subsection{Construction of single-valued elliptic polylogarithms}\label{sec:constsv}
Recall that the fundamental group of the punctured torus $E_\tau^\times$ can be described as 
\begin{equation}\label{eqn:pi1}
	\pi_1(E_\tau^\times) =  \langle  A,B,C \mid ABA^{-1}B^{-1}=C\rangle  \, ,
\end{equation}
where the generators $A,B$ correspond to the homotopy classes of the non-contractible cycles $\acyc$ and $\bcyc$ arising due to the non-trivial topology of the torus, whereas $C$ corresponds to the homotopy class of a small non-contractible loop $\mathfrak{C}$ around the puncture. The relation between the generators reflects that a path once around the parallelogram is contractible to a small loop $\mathfrak{C}$ around the puncture as depicted in \figref{fig:associatorrelation}.
\begin{figure}
	\centering
	\mpostuse{associatorrelation}
	\caption{Graphical representation of the cycle relation in \protect\eqn{eqn:pi1}. For better readability, we shifted the puncture away from $0$.}
	\label{fig:associatorrelation}
\end{figure}
One now has
\begin{equation}
	\pi_1(E_\tau^\times) \simeq \langle \cA,\cB\rangle \, ,
\end{equation}
which is the free group on two generators\footnote{The notation $\cA$ and $\cB$ for the generators of $\pi_1(E_\tau^\times)$ is suggestive to make contact with the generators $a$ and $b$ of the free Lie-algebra $\alg{fl}:=\Lie(a,b)\,{\subset}\,\alg{f}_2(\ellalph)$. This makes sense as there exists a group homomorphism $\pi_1(E_\tau^\times)\,{\rightarrow}\, \exp(\widehat{\alg{fl}})$, where $\widehat{\alg{fl}}\subset\widehat{\alg{f}_2(\ellalph)}$ and the hat denotes the (degree-)completion. This induces an isomorphism of the (complex) Lie-algebras $\Lie\!\left(\pi_1(E_\tau^\times)\right)\!{\simeq}\, \widehat{\alg{fl}}$, where $\Lie\!\left(\pi_1(E_\tau^\times)\right)$ is understood to be the Lie-algebra of the (pro-unipotent) completion of $\pi_1(E_\tau^\times)$ (see e.g.~\cite{EnriquezHigher}). Therefore, we can regard the generators $a$ and $b$ of $\alg{fl}\subset\alg{f}_2(\ellalph)$ as infinitesimal versions of the group generators $\cA$ and $\cB$, hence the notation. Throughout the article, we will by abuse of notation omit the hat and implicitly assume $\alg{f}_2(\ellalph)$ to be completed by the degree.} $\cA$ and $\cB$. To see this, notice that $A\mapsto \cA$, $B\mapsto \cB$ and $C\mapsto \cA\cB\cA^{-1}\cB^{-1}$ defines a group isomorphism. In other words, the non-trivial relation in \eqn{eqn:pi1} allows to eliminate the generator $C$, thereby reflecting that its corresponding path is homotopic to a combination of the paths $\acyc$ and $\bcyc$.

In analogy to genus zero, we want to describe the monodromies of the generating function $\Gamma(z|\tau)$ when analytically continuing\footnote{Since the coefficients of the elliptic KZB integral are homotopy invariant, it is irrelevant which representative of the homotopy classes we choose for analytic continuation of the solution. In other words, homotopy invariance (together with the uniqueness statement \eqref{eqn:KZB2ndsol} for solutions of the KZB equation) ensures that the map $\pi_1(E_\tau^\times)\rightarrow\alg{f}_2(\ellalph)$ given by analytic continuation is well-defined.} around $A$, $B$ and $C$, which generate $\pi_1(E_\tau^\times)$ according to \eqn{eqn:pi1}. This essentially follows by definition as the non-trivial cycles are exactly implemented as $z\,{\rightarrow}\,z\,{+}\,1$ for the $\acyc$-cycle and $z\,{\rightarrow}\,z\,{+}\,\tau$ for the $\bcyc$-cycle. From this observation, we find that the monodromies are exactly given by the A- and B-associator (cf.~\eqn{eqn:ellass}), i.e.
\begin{subequations}\label{eqn:monodromiesgenus1}
	\begin{align}
		\Gamma(z+\acyc|\tau) &= \Gamma(z|\tau)A(\tau) \, , \\
		\Gamma(z+\!\bcyc|\tau) &= \Gamma(z|\tau)B(\tau) \, , \\
		\Gamma(z + \mathfrak{C}|\tau) &= \Gamma(z|\tau)\exp\!\left(-2\pi\iunit [b,a]\right) \, ,
	\end{align}
\end{subequations}
where the last equation describes the monodromy around the small loop $\mathfrak{C}$ around the puncture corresponding to the generator $C$ of $\pi_1(E_\tau^\times)$, which follows from the asymptotics \eqref{eqn:asymp}.

With the above description of the monodromies associated to the eMPLs, we can now formulate an ansatz for a generating function for sveMPLs. Inspired by Brown's construction of svMPLs at genus zero, discussed in \secref{sec:svmpls}, and with the goal of cancelling the monodromies given in \eqn{eqn:monodromiesgenus1}, we write the \emph{generating function of single-valued eMPLs (sveMPLs)}, valued in $\alg{f}_2(\ellalph)$, as
\begin{equation}\label{eqn:esvansatz}
	\mathbf{\Gamma}(z|\tau) = \Gamma_{\ellalph}(z|\tau)\,\reverse\!\left(\overline{\Gamma_{\ellalph'}(z|\tau)}\right) \, ,
\end{equation}
where $\reverse$ denotes some anti-homomorphism\footnote{I.e.~a linear map $\reverse\colon\alg{f}_2(\ellalph)\,{\to}\,\alg{f}_2(\ellalph)$ satisfying $\reverse(w_1w_2)\texteq \reverse(w_2)\reverse(w_1)$ for $w_1,w_2\,{\in}\,\alg{f}_2(\ellalph)$. We will comment below on which maps $\reverse$ are suitable here.} and the subscripts $\ellalph$ and $\ellalph'$ denote the different sets of generators $\{a,b\}$ and $\{a',b'\}$ of the same algebra $\alg{f}_2(\ellalph)$.

As a small remark, notice that the discussion around \eqn{eqn:KZB2ndsol} showed that any solution to the KZB equation \eqref{eqn:KZBeq} is related to the solution $\Gamma(z|\tau)$ through multiplication by a function $S(\tau)$, which is constant in $z$ and $\zb$. Thus, when constructing a non-trivial single-valued analogue of $\Gamma(z|\tau)$, which cancels the monodromies \eqref{eqn:monodromiesgenus1}, it will not satisfy the KZB equation.

Using \eqn{eqn:monodromiesgenus1}, we can now compute the monodromies of $\mathbf{\Gamma}(z|\tau)$ and obtain
\begin{subequations}
	\begin{align}
		\mathbf{\Gamma}(z+\acyc|\tau) &= \Gamma_{\ellalph}(z|\tau)A_{\ellalph}(\tau)\,\reverse\!\left(\overline{A_{\ellalph'}(\tau)}\right)\reverse\!\left(\overline{\Gamma_{\ellalph'}(z|\tau)}\right) \, , \\
		\mathbf{\Gamma}(z+\bcyc|\tau) &= \Gamma_{\ellalph}(z|\tau)B_{\ellalph}(\tau)\,\reverse\!\left(\overline{B_{\ellalph'}(\tau)}\right)\reverse\!\left(\overline{\Gamma_{\ellalph'}(z|\tau)}\right) \, , \\
		\mathbf{\Gamma}(z + \mathfrak{C}|\tau) &= \Gamma_{\ellalph}(z|\tau)\exp\!\left(-2\pi\iunit [b,a]\right)\reverse\!\left(\exp\!\left(2\pi\iunit [b',a']\right)\right)\reverse\!\left(\overline{\Gamma_{\ellalph'}(z|\tau)}\right) \, .
	\end{align}
\end{subequations}
Accordingly, the conditions for the ansatz \eqref{eqn:esvansatz} to be single-valued read
\begin{subequations}\label{eqn:svgenusone}
	\begin{align}
		1 &= A_{\ellalph}(\tau)\,\reverse\!\left(\overline{A_{\ellalph'}(\tau)}\right)  \, , \label{eqn:svA}\\
		1 &= B_{\ellalph}(\tau)\,\reverse\!\left(\overline{B_{\ellalph'}(\tau)}\right)\, , \label{eqn:svB}\\
		1 &= \exp\!\left(-2\pi\iunit [b,a]\right)\reverse\!\left(\exp\!\left(2\pi\iunit [b',a']\right)\right) \, ,\label{eqn:svC}
	\end{align}
\end{subequations}
which we will call the \emph{elliptic single-valued conditions}. These equations are the genus-one version of \eqn{eqn:svgenuszero}. The corresponding geometric picture illustrating the parallel transport of the elliptic associators is depicted in \figref{fig:ellipticsvcondition}.
\begin{figure}[t]
	\centering
	\mpostuse{ellipticsvcondition}
	\caption{A- and B-cycle single-valued conditions \protect\eqref{eqn:svgenusone} sketched on the fundamental parallelogram (left) and on the torus (right).}
	\label{fig:ellipticsvcondition}
\end{figure}

Finally, we want to argue that it suffices to consider the elliptic single-valued conditions given in \eqn{eqn:svA} and \eqn{eqn:svB}. To see this, note that the non-trivial relation in the fundamental group (cf.~\eqn{eqn:pi1}) has its corresponding analogue
\begin{equation}\label{eqn:assrel}
	A(\tau)B(\tau)A^{-1}(\tau)B^{-1}(\tau) = \exp\!\left(2\pi\iunit [b,a]\right)
\end{equation}
at the level of the A- and B-associator \cite{Enriquez:EllAss}. This relation, together with the single-valued conditions \eqref{eqn:svA} and \eqref{eqn:svB} for A- and B-associator imply
\begin{equation}\label{eqn:expC}
	\begin{aligned}
		\exp\!\left(2\pi\iunit [b,a]\right) &= A_{\ellalph}(\tau)B_{\ellalph}(\tau)A_{\ellalph}^{-1}(\tau)B_{\ellalph}^{-1}(\tau) \\
		&= \reverse\!\left(\overline{A_{\ellalph'}^{-1}(\tau)}\right)\reverse\!\left(\overline{B_{\ellalph'}^{-1}(\tau)}\right)\reverse\!\left(\overline{A_{\ellalph'}(\tau)}\right)\reverse\!\left(\overline{B_{\ellalph'}(\tau)}\right) \\
		&= \reverse\!\left(\overline{B_{\ellalph'}(\tau)}\,\,\overline{A_{\ellalph'}(\tau)}\,\,\overline{B_{\ellalph'}^{-1}(\tau)}\,\,\overline{A_{\ellalph'}^{-1}(\tau)}\right) \\
		&= \reverse\!\left(\!\left(\overline{A_{\ellalph'}(\tau)B_{\ellalph'}(\tau)A_{\ellalph'}^{-1}(\tau)B_{\ellalph'}^{-1}(\tau)}\right)^{-1}\right) \\
		&= \reverse\!\left(\!\left(\overline{\exp\!\left(2\pi\iunit [b',a']\right)}\right)^{-1}\right) \\
		&= \reverse\!\left(\exp\!\left(2\pi\iunit [b',a']\right)\right) \, .
	\end{aligned}
\end{equation}
Here we have assumed that the inverse commutes with $\reverse$, where we will comment in more detail below how this affects the possible choices of $\reverse$. Furthermore, we used relation \eqref{eqn:assrel} for the A- and B-associator in both alphabets. From the above calculation, we can now immediately conclude that the relation \eqref{eqn:assrel}, together with the single-valued conditions \eqref{eqn:svA} and \eqref{eqn:svB}, implies the single-valued condition for the generator $C$ given in \eqn{eqn:svC}. This exactly corresponds to the isomorphic presentations of $\pi_1(E_\tau^\times)$ and the elimination of the generator $C$ described above.

\paragraph{The anti-homomorphism $\reverse$.}
The map $\reverse\colon\alg{f}_2(\ellalph)\,{\to}\,\alg{f}_2(\ellalph)$ above was only specified to be an anti-homomorphism, without giving an exact definition. It is necessary to have an analogue of the word-reversal operation $\ \widetilde{}\ $ from genus zero (cf.~\eqn{eqn:GG}). However, there is in principle some ambiguity in the choice of this map at genus one: in order to obtain single-valued elliptic polylogarithms, i.e.~doubly-periodic functions on the torus without branch cuts, the anti-homomorphism $\reverse$ is in addition required to \emph{commute with taking the inverse} of an elliptic associator\footnote{Since the elliptic associators are group-like w.r.t.~the Hopf algebra structure introduced in \secref{paragraph:Hopf}, taking the inverse coincides with the application of the antipode $\varsigma$. This can be deduced directly from the Hopf algebra axioms.} as we used already in \eqn{eqn:expC} and $\reverse$ should not be the inverse (i.e.~the antipode) itself. Furthermore, $\reverse$ should \emph{commute with the coproduct} $\Delta$ (cf.~\secref{paragraph:Hopf}) as well. Apart from that, being an analogue of the reversal map of genus zero, $\reverse$ should \emph{respect the length of the words}. In summary, the map $\reverse$ should be an \emph{anti-homomorphism of the graded Hopf algebra} $\alg{f}_2(\ellalph)$.

Matching the above conditions, two natural choices include: the mirror map\footnote{The name mirror map (also used in \rcite{Brownhyperlogs}) is motivated by the fact that it reverts (or \emph{mirrors}) the order of a given word in $\alg{f}_2(\ellalph)$, which follows as the map acts as the identity on the generators and is defined to be an antihomomorphism.} $\reverse(a)\texteq a$ and $\reverse(b)\texteq b$ as well as $\reverse(a)\texteq a$ and $\reverse(b)\texteq{-}\,b$, which essentially corresponds to the requirement of preserving the (nested) commutators $\ad^k(b)(a)$ for $k\,{\geq}\,0$. These two choices for $\reverse$ are not equivalent, since swapping order of letters $a,b$ in $\ad^k(b)(a)$ results in a factor of $(-1)^k$, but both are valid options for $\reverse$. In the following, we will choose $\reverse$ to act as the identity on the generators $a$ and $b$, i.e.
\begin{equation}\label{eqn:Rab}
	\reverse(a)=a,\quad \reverse(b)=b\quad\implies\quad \reverse\!\left(\ad^k(b)(a)\right)=(-1)^k\ad^k(b)(a).
\end{equation}
This choice for $\reverse$ will proof useful and natural when studying the degeneration of the single-valued condition \eqref{eqn:svA} and connecting it to the genus-zero condition \eqref{eqn:svgenuszero}. We will speculate on other possibilities of $\reverse$ in \secref{sec:openqs}.

\paragraph{On the single-valued KZB equation.}
As pointed out, the function $\mathbf{\Gamma}(z|\tau)$ constructed above is \emph{not} a solution to the KZB equation. Instead we find,
\begin{equation}\label{eqn:svODE}
	\dd\mathbf{\Gamma}(z|\tau)=-\omega_{\raisebox{-1.2pt}{\scriptsize$\mathrm{KZB},\ellalph$}}(z|\tau)\,\mathbf{\Gamma}(z|\tau)-\mathbf{\Gamma}(z|\tau)\,\reverse\!\left(\overline{\omega_{\raisebox{-1.2pt}{\scriptsize$\mathrm{KZB},\ellalph'$}}(z|\tau)}\right),
\end{equation}
where we denote the respective alphabets in which the KZB form is expanded as subscripts. We can furthermore analyze the asymptotic of $\mathbf{\Gamma}(z|\tau)$ for $z\,{\to}\,0$: using the map $\reverse$ as given in \eqn{eqn:Rab} and the asymptotic of $\Gamma(z|\tau)$ from \eqn{eqn:asymp}, we find
\begin{equation}\label{eqn:svasymp}
	\mathbf{\Gamma}(z|\tau)\sim|2\pi\iunit z|^{-2[b,a]}\quad\text{as}\quad z\to0.
\end{equation}
That means, we constructed the (unique) solution to \eqn{eqn:svODE} with the asymptotic from \eqn{eqn:svasymp}.

\paragraph{Solving the single-valued condition.}
It now remains to solve the single-valued conditions \eqref{eqn:svgenusone}, where it follows from a result of Brown \cite{Brownhyperlogs} that the solution exists and is unique. Since this result is central for the construction of single-valued eMPLs, we repeat the details and proof of this result in \appref{app:svol} for convenience of the reader.
In order to construct an explicit solution, we start by writing an ansatz for the second set of generators $a'$ and $b'$ in terms of $a$ and $b$, with which we then recursively solve the single-valued conditions \eqref{eqn:svgenusone} for the unconstrained parameters in this ansatz. For the generic ansatz
\begin{equation}\label{eqn:abexpansion}
	\begin{aligned}
		a' = \sum_{w\in \ellalph^\times} \alpha_w\,w \, , \qquad	b' = \sum_{w\in \ellalph^\times} \beta_w\,w \, ,
	\end{aligned}
\end{equation}
we obtain from \eqn{eqn:svgenusone} the relations
\begin{subequations}
	\begin{align}
		0 &= \coeffA(0|\tau)a + \overline{\coeffA(0|\tau)}(\alpha_a\reverse(a)+\alpha_b\reverse(b)) + \ldots\notag\\
		&=-a -(\alpha_a\reverse(a)+\alpha_b\reverse(b)) + \ldots 	\, , \\
		0 &=\coeffB(-1|\tau)b + \coeffB(0|\tau)a + \overline{\coeffB(-1|\tau)}(\beta_a\reverse(a)+\beta_b\reverse(b)) + \overline{\coeffB(0|\tau)}(\alpha_a\reverse(a)+\alpha_b\reverse(b)) + \ldots\notag\\
		&=2\pi\iunit\, b -\tau\,a -2\pi\iunit(\beta_a\reverse(a)+\beta_b\reverse(b)) -\overline{\tau}(\alpha_a\reverse(a)+\alpha_b\reverse(b)) + \ldots\,,
	\end{align}
\end{subequations}
to first order in the length of the words, where we used the expression for the coefficients $\coeffA$ and $\coeffB$ from \eqn{eqn:assexpansion}. Choosing the map $\reverse$ as defined in \eqn{eqn:Rab} results in the system of equations
\begin{subequations}
	\begin{align}
		0 &= -1-\alpha_a \, , &&0 = -\alpha_b\, , \\
		0 &= 2\pi\iunit - 2\pi\iunit\beta_b - \overline{\tau}\,\alpha_b \, , &&0= -\tau -2\pi\iunit\,\beta_a - \overline{\tau}\,\alpha_a \, ,
	\end{align}
\end{subequations}
which has the (unique) solution
\begin{equation}
	\alpha_a = -1 \, ,\qquad \alpha_b = 0\, , \qquad \beta_a = -\frac{\Im(\tau)}{\pi}\, ,\qquad \beta_b = 1 \, .
\end{equation}
Thus, we can write to first order
\begin{equation}\label{eqn:apbp}
	a' = -a + \cO(|w|\,{\geq}\,3)\, , \qquad b' = -\frac{\Im(\tau)}{\pi}a +b+ \cO(|w|\,{\geq}\,3) \,,
\end{equation}
where the corrections at wordlength two vanish and the next corrections to $a'$ and $b'$ are of wordlength three in the letters $a$ and $b$, which we denote by the notation $\cO(|w|\,{\geq}\,3)$. These next terms are shown in \appref{app:apbp}. An analogous change of alphabet was formulated in terms of zeta generators in \rcite{SST}.

We then write the generating series of sveMPLs as
\begin{align}\label{eqn:svGexpand}
	\mathbf{\Gamma}(z|\tau)&=\sum_{w\in \ellalph^\times}\svcoeff{w}{z}\,w=1+\left(-2\Re\!\left(\empltg{1}{z}\right)+2\pi\frac{\Im(z)^2}{\Im(\tau)}\right)[b,a] + \cO(|w|\geq 3),
\end{align}
where we denote the single-valued coefficients for a word $w$ in letters $a,b$ by $\svcoeff{w}{z}$ and used the explicit formul\ae{} of some eMPLs \eqref{eqn:explicitInts} to cancel terms from the second to third line. This shows that the first non-trivial example of a single-valued eMPL appears at length $2$. We will take up this example in more detail in \secref{sec:ellipticlog}. 

Note that all other coefficients $\svcoeff{w}{z}$ for words $w$ up to word length two vanish. This is indeed expected since, in the single-valued construction, the coefficients $\alpha_{a^n},\,\alpha_{b^n},\,\beta_{a^n}$ and $\beta_{b^n}$ all seem to vanish for $n\,{\geq}\,2$ in \eqn{eqn:abexpansion}. The corrections to $a'$ and $b'$ at length one, together with the differential equation \eqref{eqn:svODE} for the generating function of sveMPLs, then directly imply the vanishing of $\svcoeff{a^k}{z}$ and $\svcoeff{b^k}{z}$ for all $k\,{\geq}\,1$.

\paragraph{Shuffle relations for sveMPLs.}

One key property of the elliptic KZB solution and eMPLs is that they satisfy the shuffle relations. Therefore, it is desirable to require the same for any construction of single-valued eMPLs. In order to show that the construction proposed in this work indeed is a reasonable framework for sveMPLs, we will now briefly elaborate on why shuffle relations are in fact preserved by our construction. The argument here is an adapted version of Brown's construction at genus zero \cite{Brownhyperlogs}.

Referring to \secref{paragraph:Hopf}, showing that the coefficients of $\mathbf{\Gamma}(z|\tau)$ satisfy the shuffle relations is equivalent to showing that the generating series $\mathbf{\Gamma}(z|\tau)$ is group-like with respect to the coproduct $\Delta$ defined in \eqn{eqn:coproduct}. To do this, notice first that the map $\reverse$ defined in \eqn{eqn:Rab} commutes with $\Delta$, which can be immediately seen from the respective definitions. Thus we can write
\begin{equation}\label{eqn:shuffle1}
	\begin{aligned}
		\Delta(\mathbf{\Gamma}(z|\tau)) =& \Delta\!\left(\Gamma_\ellalph(z|\tau)\,\reverse\!\left(\overline{\Gamma_{\ellalph'}(z|\tau)}\right)\right) \\
		=&\left(\Gamma_\ellalph(z|\tau)\otimes \Gamma_\ellalph(z|\tau)\right)(\reverse\otimes\reverse)\,\overline{\Delta\left(\Gamma_{\ellalph'}(z|\tau)\right)} \, ,
	\end{aligned}
\end{equation}
where we have used that $\Gamma(z|\tau)$ is group-like (cf.~\secref{paragraph:Hopf}). 

Above \eqn{eqn:ellass}, it was discussed that the elliptic associators are group-like series as well. Using commutation of $\reverse$ with the coproduct again it is obvious that the modified series $\reverse\big(\overline{A_\ellalph(\tau)}\big)$ and $\reverse\big(\overline{B_\ellalph(\tau)}\big)$ appearing in the elliptic single-valued conditions\footnote{Notice that compared to the elliptic single-valued conditions, we currently consider these series as expanded in the set of generators $\ellalph$ instead of $\ellalph'$, hence the subscript.} \eqref{eqn:svgenusone} are likewise group-like. As detailed in \appref{app:svol}, we can describe $\ellalph'$, i.e.~the solution of the single-valued condition \eqref{eqn:apbp} constructed above, by a map $\rho:\alg{f}_2(\ellalph)\,{\rightarrow}\,\alg{f}_2(\ellalph)$ such that $\ellalph'\texteq\rho(\ellalph)\texteq\{\rho(a),\rho(b)\}$ and $\rho$ commutes with $\Delta$. This now immediately implies that $\Gamma_{\ellalph'}(z|\tau)$ is group-like as 
\begin{equation}\label{eqn:shuffle2}
	\Delta(\Gamma_{\ellalph'}(z|\tau))=\Delta(\rho(\Gamma_{\ellalph}(z|\tau)))=(\rho\otimes\rho)(\Gamma_{\ellalph}(z|\tau)\otimes\Gamma_\ellalph(z|\tau))=\Gamma_{\ellalph'}(z|\tau)\otimes\Gamma_{\ellalph'}(z|\tau) \, ,
\end{equation}
where we have only used that $\rho$ and $\Delta$ commute as well as the fact that $\Gamma_\ellalph(z|\tau)$ is group-like. Putting together \eqns{eqn:shuffle1}{eqn:shuffle2}, we can conclude that indeed
\begin{equation}\label{eqn:shuffle3}
	\begin{aligned}
		\Delta(\mathbf{\Gamma}(z|\tau)) =&\left(\Gamma_\ellalph(z|\tau)\otimes \Gamma_\ellalph(z|\tau)\right)(\reverse\otimes\reverse)\overline{\Delta\left(\Gamma_{\ellalph'}(z|\tau)\right)} \\
		=&\left(\Gamma_\ellalph(z|\tau)\otimes \Gamma_\ellalph(z|\tau)\right)(\reverse\otimes\reverse)(\overline{\Gamma_{\ellalph'}(z|\tau)}\otimes\overline{\Gamma_{\ellalph'}(z|\tau)}) \\
		=&\mathbf{\Gamma}(z|\tau)\otimes \mathbf{\Gamma}(z|\tau) \, ,
	\end{aligned}
\end{equation}
which shows that the generating function of sveMPLs is indeed group-like. This finally implies that its coefficients, i.e.~the sveMPLs, satisfy the shuffle relations.

\subsection{Degeneration of the single-valued condition} \label{sec:degen}

In the degeneration limit $\tau\,{\to}\,\iunit\infty$, the elliptic A-associator becomes \cite{CEE,Enriquez:EllAss,Enriquez:Emzv}
\begin{align}
	A^{(\infty)}:=\lim_{\tau\to\iunit\infty}A(\tau)=\Drinfeld(\mathrm{Ber}_b(a),t)e^{2\pi\iunit\mathrm{Ber}_b(a)}\Drinfeld(\mathrm{Ber}_b(a),t)^{-1}
\end{align}
where\footnote{This notation refers to $\mathrm{Ber}_b(a)$ essentially being the generating series of Bernoulli numbers, which was also used in \rcite{schneps2022fayrelationssatisfiedelliptic}.} $\mathrm{Ber}_b(a)\texteq{-}\frac{\ad(b)}{e^{2\pi\iunit\ad(b)}-1}(a)$ and $t\texteq{-}\ad(b)(a)$. 
Taking again the reversal map $\reverse$ to be the anti-automorphism on $\alg{f}_2(\ellalph)$ defined by $\reverse(a)\texteq a$, $\reverse(b)\texteq b$, we can calculate the reversal $\reverse(t)\texteq{-}\,t$ and the reversal plus complex conjugation $\reverse\big(\overline{\mathrm{Ber}_b(a)}\big)\texteq{-}\,\mathrm{Ber}_b(a)$. Using this, the elliptic single-valued condition for the A-associator \eqref{eqn:svA} reduces under degeneration to\footnote{Recall that the Drinfeld associator has only real coefficients, hence $\overline{\Drinfeld}\texteq\Drinfeld$.}
\begin{equation}
\begin{aligned}
	1&=\lim_{\tau\to\iunit\infty}A_{\ellalph}(\tau)\,\reverse\!\left(\overline{A_{\ellalph'}(\tau)}\right)=A^{(\infty)}_{\ellalph}\,\reverse\!\left(\overline{A^{(\infty)}_{\ellalph'}}\right)\\
	&=\Drinfeld(\mathrm{Ber}_b(a),t)e^{2\pi\iunit \mathrm{Ber}_b(a)}\Drinfeld(\mathrm{Ber}_b(a),t)^{-1}\,\reverse\!\left(\overline{\Drinfeld(\mathrm{Ber}_{b'}(a'),t')e^{2\pi\iunit \mathrm{Ber}_{b'}(a')}\Drinfeld(\mathrm{Ber}_{b'}(a'),t')^{-1}}\right)\\
	&=\Drinfeld(\mathrm{Ber}_b(a),t)e^{2\pi\iunit \mathrm{Ber}_b(a)}\Drinfeld(\mathrm{Ber}_b(a),t)^{-1}\widetilde{\Drinfeld}(-\mathrm{Ber}_{b'}(a'),-t')^{-1}e^{2\pi\iunit \mathrm{Ber}_{b'}(a')}\widetilde{\Drinfeld}(-\mathrm{Ber}_{b'}(a'),-t')
\end{aligned}
\end{equation}
Using the property $\Drinfeld(x,y)^{-1}\texteq\Drinfeld(y,x)$ of the Drinfeld associator and identifying $\szero\texteq t$, $\sone\texteq \mathrm{Ber}_b(a)$, $\szero'\texteq{-}\,t'$ and $\sone'\texteq {-}\,\mathrm{Ber}_{b'}(a')$, this translates into
\begin{equation}
	\Drinfeld(\szero,\sone)^{-1}e^{2\pi\iunit \sone}\Drinfeld(\szero,\sone)=\widetilde{\Drinfeld}(\szero',\sone')e^{2\pi\iunit \sone'}\widetilde{\Drinfeld}(\szero',\sone')^{-1},
\end{equation}
which is precisely the genus-zero single-valued condition for svMPLs \eqref{eqn:svgenuszero} from \rcite{BrownSVMPL}. This implies, using $\reverse$ as in \eqn{eqn:Rab}, by solving the elliptic single-valued condition \eqref{eqn:svgenusone}, we simultaneously solve the genus-zero single-valued condition \eqref{eqn:svgenuszero}.


\section{Examples} \label{sec:examples}
\subsection{The single-valued elliptic logarithm}\label{sec:ellipticlog}
\begin{figure}[t]
	\centering
	\begin{subfigure}[t]{0.3\textwidth}
		\centering
		\includegraphics[width=\textwidth]{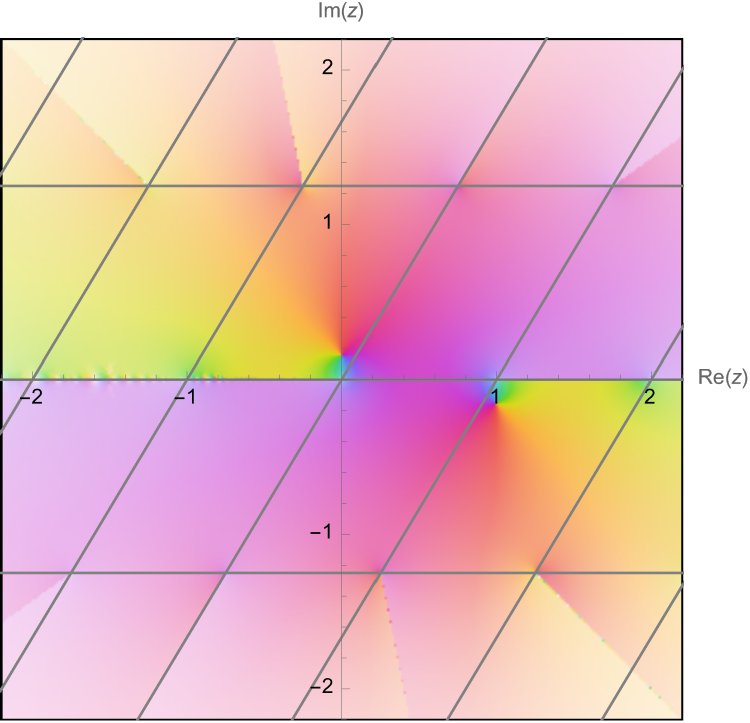}
		\caption{}
	\end{subfigure}
	\hspace{-4ex}
	\begin{subfigure}[t]{0.15\textwidth}
		\centering
		\includegraphics[width=0.5\textwidth]{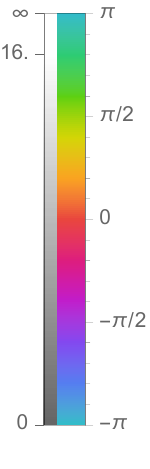}
	\end{subfigure}
	\begin{subfigure}[t]{0.4\textwidth}
		\centering
		\includegraphics[width=\textwidth]{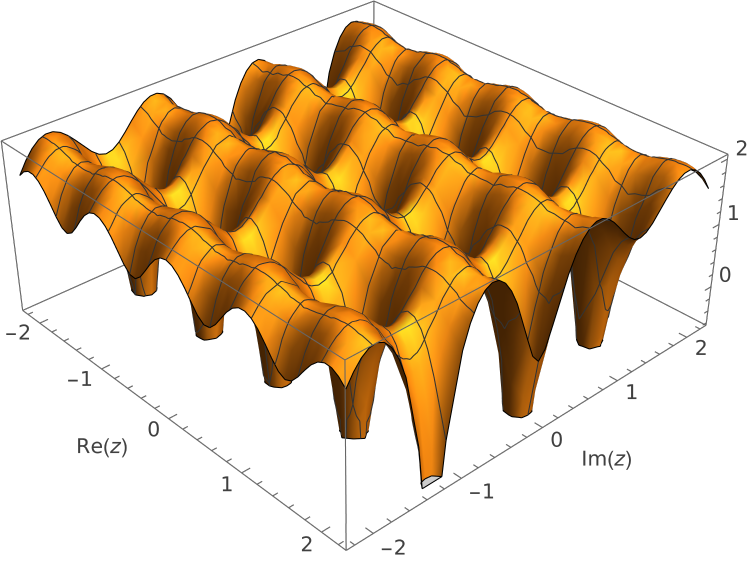}
		\caption{}
	\end{subfigure}
	\caption{(a) complex plot of $\tilde{\Gamma}(1;z|\tau)$, and (b) 3D plot of $\tilde{\Gamma}^{\sv}(1;z|\tau)\texteq\svcoeff{ab}{z}$, both for $\tau\texteq\tfrac{3}{4}\,{+}\,\tfrac{5}{4}\iunit$. In (a) we can see the branch cuts and thus the multi-valued nature of $\tilde{\Gamma}(1;z|\tau)$, while in (b) all branch cuts have been canceled and we are left with the single-valued real function $\tilde{\Gamma}^{\sv}(1;z|\tau)$, which is doubly-periodic.}
	\label{fig:G1}
\end{figure}
The elliptic analog of the logarithm $G(1;z)\texteq{-}\log(1\,{-}\,z)$ is given by
\begin{equation}
	\tilde\Gamma(1;z|\tau)=\int_0^z\gkern{1}(t|\tau)\,\dd t=\log(\theta(z|\tau))-\log(\theta'(0|\tau))+\log(-2\pi\iunit),
\end{equation}
where the integral was regularized at the lower integration boundary. Due to the logarithm, this function has a branch cut (along the negative imaginary axis) and is neither periodic along the $\acyc$-cycle nor the $\bcyc$-cycle. For the monodromies around the cycles, we find explicitly 
\begin{subequations}\label{eqn:G1monodromy}
\begin{align}
	\tilde\Gamma(1;z+1|\tau)&=\tilde\Gamma(1;z|\tau)-\pi\iunit,\\
	\tilde\Gamma(1;z+\tau|\tau)&=\tilde\Gamma(1;z|\tau)+\pi\iunit-\pi\iunit\tau-2\pi\iunit z.
\end{align}
\end{subequations}
From \eqn{eqn:fkern}, we can immediately conclude that
\begin{equation}
	\tilde\Gamma(1;z|\tau) = \int_0^z\fkern{1}(t|\tau)\,\dd t - \frac{2\pi\iunit}{\Im(\tau)}\int_0^z \Im(t)\,\dd t = \emplcoeff{ab}{z},
\end{equation}
and therefore the single-valued construction yields
\begin{equation}
	\begin{aligned}
		\tilde\Gamma^{\mathrm{sv}}(1;z|\tau) &= \svcoeff{ab}{z}= 2\Re\!\left(\empltg{1}{z}\right)-\frac{2\pi}{\Im(\tau)}\Im(z)^2 \, ,
	\end{aligned}
\end{equation}
where we have used the explicit formulas \eqref{eqn:explicitInts}. Clearly, in this combination the branch cuts cancel, since we now have $\log|1-e^{2\pi\iunit z}|^2$, and the function becomes doubly periodic on the torus, as shown in \figref{fig:G1}.

Similarly, we can relate
\begin{equation}
	\empltg{k}{z}=(-1)^{k+1}\emplcoeff{ab^k}{z}
\end{equation}
for $k\,{\geq}\,0$, which immediately yields a single-valued version $\tilde\Gamma^{\mathrm{sv}}(k;z|\tau)\texteq(-1)^{k+1}\svcoeff{ab^k}{z}$ of all these eMPLs via the construction presented in this work. Conjecturally, we can extend this to all eMPLs, but this will be beyond the scope of the current project.


\subsection{An example of depth three}

As a non-trivial example, where also the corrections to $a'$ and $b'$ from \eqn{eqn:apbp} contribute, let us have a look at the depth-three case related to the word $b^2a$ in the expansion \eqref{eqn:svGexpand} of the generating series of sveMPLs, i.e.~the coefficient $\emplcoeff{b^2a}{z}$.

\begin{figure}[t]
	\centering
	\begin{subfigure}[t]{0.3\textwidth}
		\centering
		\includegraphics[width=\textwidth]{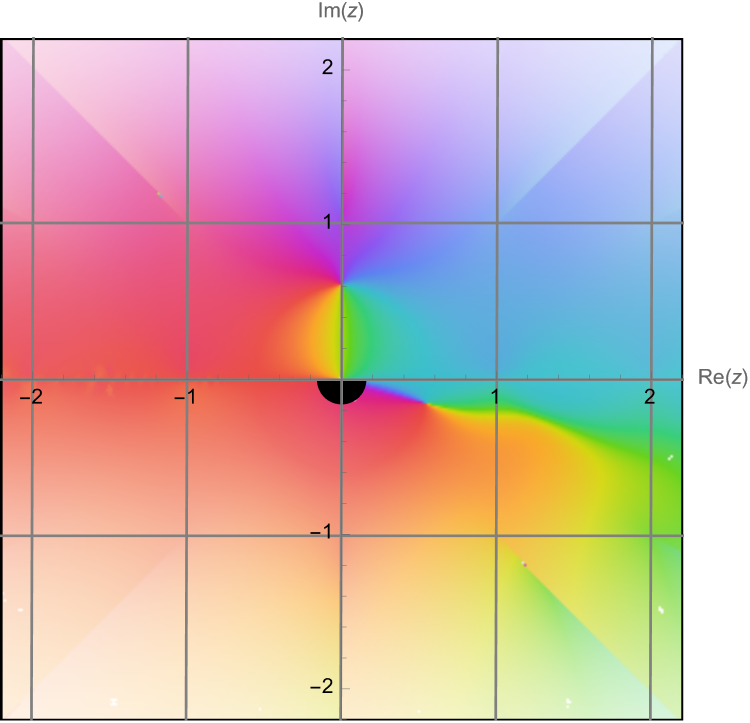}
		\caption{}
	\end{subfigure}
	\hspace{-4ex}
	\begin{subfigure}[t]{0.15\textwidth}
		\centering
		\includegraphics[width=0.5\textwidth]{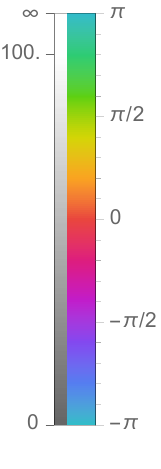}
	\end{subfigure}
	\begin{subfigure}[t]{0.4\textwidth}
		\centering
		\includegraphics[width=\textwidth]{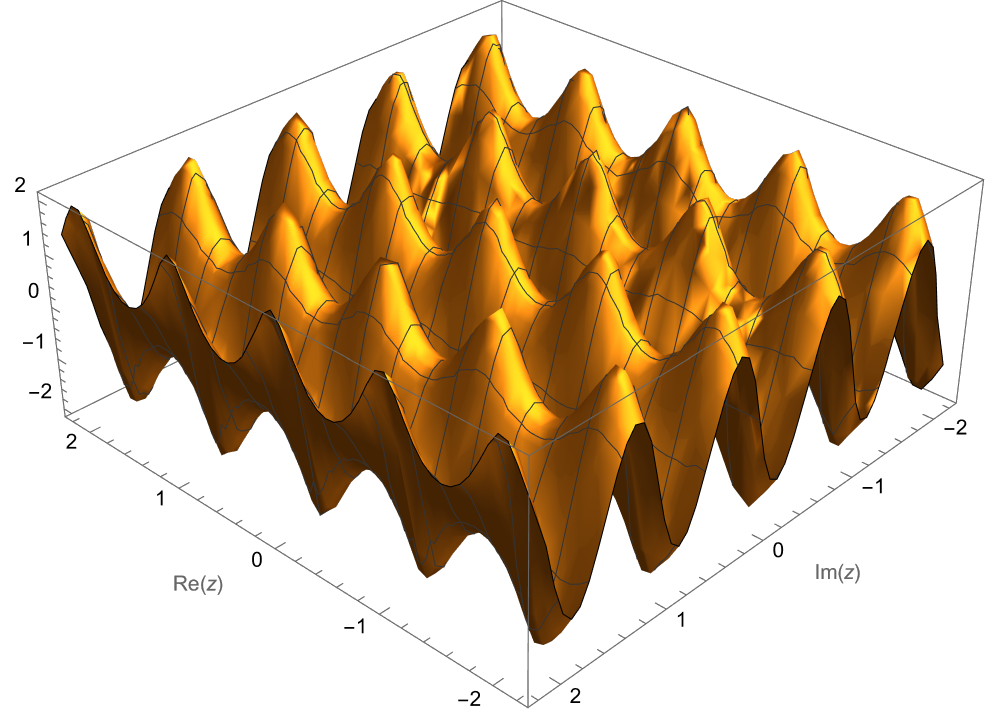}
		\caption{}
	\end{subfigure}
	\caption{Plots showing (a) the multi-valued function $\emplcoeff{b^2a}{z}$ and (b) its single-valued analogue $\svcoeff{b^2a}{z}$ given in \protect\eqn{eqn:d3ex}, both for $\tau\texteq0.003{+} 1.01 \iunit$. Numerical difficulties around $z\texteq0$ have been cut out of the plots.}
	\label{fig:d3ex}
\end{figure}

The expansion yields the expression
\begin{equation}\label{eqn:d3ex}
	\begin{aligned}
		\svcoeff{b^2a}{z}=&\,\emplcoeff{b^2a}{z}-\overline{\emplcoeff{ab^2}{z}}-\emplcoeff{b^2}{z}\left(\overline{\emplcoeff{a}{z}}+\frac{\Im(\tau)}{\pi}\overline{\emplcoeff{b}{z}}\right)\\
		&-\emplcoeff{b}{z}\left(\overline{\emplcoeff{ab}{z}}+\frac{\Im(\tau)}{\pi}\overline{\emplcoeff{b^2}{z}}\right)\\
		&-\frac2\pi\left(\zeta_2\Im(\tau)+\pi\Re\!\left(\coeffA(0,1|\tau)\right)\right)\overline{\emplcoeff{b}{z}}\\
		=&-2\iunit\Im(\empltg{2}{z})-\frac{4\iunit\pi \Im(z)}{\Im(\tau)}\Re\!\left(\empltg{1}{z}\right)\\
		&+\frac{4\iunit\pi\Im(z)}{\Im(\tau)}\left(\frac{\pi\Im(z)}{\Im(\tau)}+\Re\left(\coeffA(0,1|\tau)\right)+\frac{\zeta_2\Im(\tau)}{\pi}\right) \, ,
	\end{aligned}
\end{equation}
where we have also made use of the relation \eqref{eqn:modrelFinal} as well as similar identities as in \eqn{eqn:explicitInts} up to length three. In \figref{fig:d3ex}, we plotted the multi-valued function $\emplcoeff{b^2a}{z}$ and its single-valued counterpart from \eqn{eqn:d3ex}.
%

\subsection{Elliptic Bloch--Wigner dilogarithm}\label{sec:ebwd}
The elliptic Bloch--Wigner dilogarithm (eBWd) $\BW^\tau$ is an elliptic generalization of the genus-zero Bloch--Wigner dilogarithm $\BW$, which in turn is a single-valued version of the dilogarithm $\Li_2$.
The eBWd is defined as \cite{Bloch}
\begin{equation}
	\BW^\tau(z)=\sum_{k\in\zZ}\BW(zq^k),\quad\text{where}\quad\BW(z)=\Im(\Li_2(z)+\log|z|\log(1-z)),
\end{equation}
and can be expressed in terms of eMPLs as \cite{Broedel_2020}
\begin{equation}
	\begin{aligned}
	\BW^\tau(z)&=\Im(\tau)\Re\!\left(\tilde{\Gamma}(2;z|\tau)\!\right)+2\pi\Re\!\left(\tilde{\Gamma}(1,0;z|\tau)\!\right)-2\pi\Re(z)\Re\!\left(\tilde{\Gamma}(1;z|\tau)\!\right)\\
	&\quad+2\Re(z)\Big(\pi\Re(\Aemzv{1,0})+\zeta_2\Im(\tau)\Big).
	\end{aligned}
\end{equation}

As for genus zero, there is also a notion of a (multi-valued) elliptic dilogarithm $\mathrm{eLi}_2$ defined as (a suitably modified) average of the dilogarithm at genus zero over the Tate curve\footnote{The Tate curve is given by $\zC/q^\zZ\,{\simeq}\, E_\tau$, where $q\texteq\!e^{2\pi\iunit\tau}$ and the identification is given by $\exp:E_\tau\,{\rightarrow}\,\zC/q^\zZ$.} put forward by Levin \cite{Levin} as well as Brown--Levin \cite{BrownLevin}.
In \rcite[Lemma 70]{BrownLevin}, $\mathrm{eLi}_2$ has been related to the notions of eMPLs studied in this work, which amounts to the equality
\begin{equation}
\begin{aligned}
	\mathrm{eLi}_2(z|\tau) = 2\pi \Big(\emplcoeff{a^2b}{z}+\emplcoeff{aba}{z}\Big) + \tau \emplcoeff{ab^2}{z} \, .
\end{aligned}
\end{equation}
So, it is natural to expect the eBWd to appear as the single-valued version of this combination and indeed, after various simplifications (including the identity \eqref{eqn:modrelFinal} proven in \appref{app:coeffid} as well as the explicit formulas \eqn{eqn:explicitInts} from \appref{app:svexp} and higher length analogues), we find precisely
\begin{equation}
	\BW^\tau(z) = 2\pi \Big(\svcoeff{a^2b}{z}+\svcoeff{aba}{z}\Big) + \Im(\tau) \svcoeff{ab^2}{z} \, .
\end{equation}
%

\section{Open question}\label{sec:openqs}

In this article, we have put forward a construction of single-valued elliptic polylogarithms based on expressing the condition of trivial monodromy for the solution to the KZB equation in terms of associators. While our formalism allows to identify single-valued elliptic polylogarithms as coefficients of an algebraic generating series and to relate those coefficients to known implementations of multi-valued polylogarithms, there are several open questions related to our construction:  

\begin{itemize}
	\item \textbf{Elliptic hyperlogarithms.} Is it possible to draw a complete analogy to Brown's genus-zero construction by developing a concise algebraic framework for elliptic polylogarithms? This should be done by defining an algebra of elliptic hyperlogarithms, study its automorphism group and prove completeness and uniqueness for homotopy invariant eMPLs. 
	Moreover, one might ask, whether it is possible to develop a formalism resulting in eMPLs with particular fixed but nontrivial monodromies around the cycles and the puncture.
	\item \textbf{Uniqueness of $\mathbf{\Gamma}(z|\tau)$.} In \secref{sec:constsv} we already discussed that there can be some ambiguity in the word reversal function used. Thus, it is important to investigate the uniqueness of our construction of sveMPLs. This includes comparing to other examples of single-valued elliptic polylogarithms.
	\item \textbf{More punctures.} Our current construction delivers single-valued elliptic polylogarithms of a single variable $z$. Is it possible to generalize the procedure to multi-variable single-valued polylogarithms, that is, to more marked points on the elliptic curve? The key ingredients for this endeavor, like the connection, are essentially already laid out in \rcite{BrownLevin}.
	\item \textbf{Single-valued construction for the full universal KZB connection.} Can our formalism be extended to the full KZB connection on the moduli space of marked points on the elliptic curve including the dependence on the geometric parameter $\tau$, using the language developed by Calaque, Enriquez and Etingof \cite{CEE}? Once available and constrained to special values of $z$, such a formalism should yield the available single-valued constructions for elliptic zeta values and iterated integrals of Eisenstein series. From a different perspective, this question is going to be answered in the upcoming publication of Schlotterer, Sohnle and Tao \cite{SST}: here the opposite direction is illuminated, as there the available single-valued construction for elliptic multiple zeta values (or iterated Eisenstein integrals) is extended towards a construction of single-valued elliptic polylogarithms. 
	\item \textbf{One-loop double copy.} Could one relate our construction to existing notions of elliptic double-copies and KLT relations for genus-one string scattering amplitudes? In particular it would be interesting to see if the analytic underpinnings in Stieberger's one-loop KLT relations \cite{Stieberger:2022lss,Stieberger:2023nol,Mazloumi:2024wys} can be explained by our approach based on the triviality of monodromies? In addition, it would be nice to relate to the (elliptic) double-copy construction put forward by~\rcite{Bhardwaj:2023vvm,Pokraka:2025zlh}.
	\item \textbf{Single-valued map for differential forms.} Is it possible to lift the single-valued construction for genus-zero string scattering amplitudes of Brown--Dupont \cite{Brown:2018omk,Brown:2019wna} to the genus-one case by finding appropriate dual differential forms for an elliptic single-valued pairing?
	\item \textbf{Higher-genus.} The present single-valued construction essentially lifts Brown's genus-zero construction to including the non-trivial A- and B-cycles symmetries of the elliptic curve. In terms of letters, i.e.~generators of the underlying algebra, each letter is associated to the symmetry induced by one non-trivial cycle. The generalization to higher-genus Riemannian manifolds of genus $g$ involves now $2g$ letters, whose relations are governed by $\mathrm{Sp}(2g,\zZ)$ transformations. Is it possible to straightforwardly generalize our construction (combinatorially) to a single-valued formalism for Riemann surfaces of arbitrary genus? Finally, once a construction of single-valued higher-genus polylogarithms has been identified, it is natural to investigate its behavior under degeneration of the underlying Riemann surface as commented upon in \secref{sec:degen} for the elliptic case.
\end{itemize}
%


\vspace*{1cm}
\noindent{\LARGE\textbf{Appendix}}
\phantomsection
\addcontentsline{toc}{section}{Appendix}
\appendix
\addtocontents{toc}{\protect\setcounter{tocdepth}{0}}

\appendixsection{\texorpdfstring{Coefficients $\coeffA$ and $\coeffB$}{Coefficients ϖA and ϖB}}
In this appendix we collect several explicit results for the coefficients $\coeffA$ and $\coeffB$ that we used in our computations, including an identity used to simplify various expressions in \secref{sec:svempls}.

\appendixsubsection{Explicit results} \label{app:coeffs}
The coefficients can be explicitly evaluated (under the assumption of using straight-line paths, which in the combinations used in this article does not matter since they are homotopy invariant) to read\footnote{Using here $\zeta_0\texteq{-}\,1/2.$}\ymnote{also added a correction here by adding $k\neq 1$ in the first line}
\begin{subequations}
	\begin{align}
		\coeffA(k|\tau)&=\begin{cases}
			2\zeta_k,&k\text{ even},\\
			0,&k\text{ odd and }k\neq1,
		\end{cases}\\
		\coeffA(k_1,k_2|\tau)&=\begin{cases}
			2\zeta_{k_1}\zeta_{k_2},& k_1,k_2,k_1+k_2\text{ even},\\
			0,&k_1+k_2\text{ even and }l_1,l_2\text{ odd},
		\end{cases}\\
		\coeffA(\underbrace{0,\ldots,0}_{n}|\tau)&=\frac{(-1)^n}{n!},\\
		\coeffA(1|\tau)&=\pi\iunit,\\
		\coeffA(1,0|\tau)&=-\pi\iunit-\coeffA(0,1|\tau),\\
		\coeffB(\underbrace{0,\ldots,0}_{n}|\tau)&=\frac{(-\tau)^{n}}{n!},\\
		\coeffB(\underbrace{-1,\ldots,-1}_{n}|\tau)&=\frac{(2\pi\iunit)^n}{n!},\\
		\coeffB(1|\tau)&=-\pi\iunit,\\
		\coeffB(-1,0|\tau)&=\coeffB(0,-1|\tau)=-\pi\iunit\tau,\\
		\coeffB(1,-1|\tau)&=-\coeffB(-1,1|\tau)+2\pi^2,\\
		\coeffB(1,0|\tau)&=\pi\iunit\tau-\coeffB(0,1|\tau),\\
		\coeffB(0,1|\tau)&=\pi\iunit\tau+\tau\coeffA(0,1|{-}\tau^{-1})-\tau\log({-}\tau^{-1}),\\
		\coeffB(-1,0,0|\tau)&=\coeffB(0,-1,0|\tau)=\coeffB(0,0,-1|\tau)=\frac{1}{3}\pi\iunit\tau^2,\\
		\coeffB(-1,-1,0|\tau)&=\coeffB(0,-1,-1|\tau)=\coeffB(-1,0,-1|\tau)=\frac{2\pi^2\tau}{3}.
	\end{align}
\end{subequations}
Furthermore, one can use shuffle relations among the coefficients to simplify further terms.

\appendixsubsection{An identity for the coefficients}\label{app:coeffid}

In order to simplify certain expression in our calculations, e.g.~for the eBWd in \secref{sec:ebwd} or for some terms in the expansion in \secref{app:svexp}, we needed the following identity between coefficients of the A-associator:
\begin{align}\label{eqn:C2}
	0=\pi|\tau|^2\Big(\Re(\coeffA(0,1|{-}\tau^{-1}))-\Re(\coeffA(0,1|\tau))+\log|\tau|\Big)-\zeta_2\big(|\tau|^2-1\big)\Im(\tau).
\end{align}
This identity can be proven using (iterated) Eisenstein integrals as follows: we start by writing \cite{Broedel:2015hia}
\begin{equation}\label{eqn:C3}
	\coeffA(0,1|-\tau^{-1}) = \omega_{\raisebox{-2.8pt}{\scriptsize$\acyc$}}(1,0|-\tau^{-1}) = -\frac{\pi\iunit}{2} - \frac{\iunit}{2\pi}\int_{\iunit\infty+\eps}^{-\tau^{-1}}\left(G_2(\tau')-2\zeta_2\right)\dd\tau' \, ,
\end{equation}
where $\eps\,{\in}\,\iunit\zR_{>0}$ indicates tangential basepoint regularization\footnote{This implicitly involves keeping only the $\eps$-finite terms and performing the limit $\eps\,{\rightarrow}\,0$ in the end.} at the cusp $\iunit\infty$ and $G_2(\tau)$ denotes the Eisenstein series of weight\footnote{The Eisenstein series of weight two is defined as the convergent sum
	 \begin{equation*}
	 	G_2(\tau)=\sum_{\substack{m,n\in\zZ\\(m,n)\neq(0,0)}}\frac{1}{(m+n\tau)^2}=\lim_{N\rightarrow\infty}\lim_{M\rightarrow\infty}\sum_{n=-N}^N\sum_{n=-M}^M\frac{1}{(m+n\tau)^2}\, .
	 \end{equation*}
	 The second equality specifies the usage of the Eisenstein summation prescription, which is required since $G_2(\tau)$ does not converge absolutely.} two. Next, we can make use of the fact that $G_2(\tau)$ is a quasi-modular form and accordingly transforms as (see e.g.~\cite{1-2-3})
\begin{equation}\label{eqn:StransfG2}
	G_2(-\tau^{-1}) = \tau^2G_2(\tau) - 2\pi\iunit \tau
\end{equation}
under a modular $S$-transformation $\tau\,{\mapsto}\,{-}\tau^{-1}$. Upon making the substitution $\sigma\texteq{-}\tau'^{-1}$, we therefore obtain for the integral in \eqn{eqn:C3}
\begin{equation}
	\begin{aligned}
		\int_{\iunit\infty+\eps}^{-\tau^{-1}}\left(G_2(\tau')-2\zeta_2\right)\dd\tau' &= \int_{\eps}^{\tau}\left(G_2(-\sigma^{-1})-2\zeta_2\right)\frac{\dd\sigma}{\sigma^2} \\
		&= \int_{\eps}^{\tau}\left(\sigma^2G_2(\sigma)-2\pi\iunit\sigma-2\zeta_2\right)\frac{\dd\sigma}{\sigma^2} \\
		&= \int_{\eps}^{\tau}\left(G_2(\sigma) - 2\zeta_2 - \frac{2\pi\iunit}\sigma-2\zeta_2\frac{1-\sigma^2}{\sigma^2}\right)\dd\sigma\, ,
	\end{aligned}
\end{equation}
where we have made use of \eqn{eqn:StransfG2}. Plugging this result into \eqn{eqn:C3}, we obtain
\begin{align}
	\coeffA(0,1|-\tau^{-1}) &= \coeffA(0,1|\tau) - \frac{\iunit}{2\pi}\int_{\eps}^{\iunit\infty+\eps}\dd\sigma\left(G_2(\sigma) - 2\zeta_2\right) - \frac{\iunit}{2\pi}\int_{\eps}^{\tau}\dd\sigma\left(- \frac{2\pi\iunit}\sigma-2\zeta_2\frac{1-\sigma^2}{\sigma^2}\right) \notag\\
	&= \coeffA(0,1|\tau) -\log(\tau) - \frac{\iunit\zeta_2}{\pi}\left(\frac{1}{\tau}+\tau\right) +\frac{\iunit\pi}{2}\, ,
\end{align}
which agrees with the formula found in ref.~\cite[eq.~(G.8)]{SST}.
Finally, taking the real part of the above equation yields
\begin{equation}\label{eqn:modrelFinal}
	\Re\!\left(\coeffA(0,1|-\tau^{-1})\right) = \Re\!\left(\coeffA(0,1|\tau)\right) -\log|\tau| - \frac{\zeta_2\Im(\tau)}{\pi|\tau|^2}\left(1-|\tau|^2\right) \, ,
\end{equation}
which immediately implies the desired result.

\appendixsection{Existence of a solution to the elliptic single-valued conditions}
\label{app:svol}

In this section, we want to argue that a unique solution to the elliptic single-valued conditions \eqns{eqn:svA}{eqn:svB} exists. Notice that this statement has been settled by Brown in \rcite{Brownhyperlogs} in the context of the single-valued construction at genus zero and essentially translates to our setting without modification. However, we repeat the proof here for convenience of the reader.

Let us start by introducing some preparatory notions. Let $\alg{f}(X)$ be the free, associative algebra generated by the finite set $X$. Consider some word $w\,{\in}\,X^\times$, for $X^\times$ the set of all words from letters in $X$ including the empty word. Then we define the map $c_w:\alg{f}(X)\,{\rightarrow}\,\zC$, which sends an element $S\,{\in}\,\alg{f}(X)$ to the coefficient of $w$ in $S$, such that $S\texteq\sum_{w\in X^\times}c_w(S)\,w$. With this, we can now state the precise result.

Let $n\,{\geq}\,0$ and consider two sets of formal power series $S_n,S_n'\,{\in}\,\alg{f}(X)$ for $X\texteq\{x_1,\ldots, x_n\}$. Moreover, suppose that
\begin{equation}\label{eqn:constterm}
	c := c_e(S_i) = c_e(S_i') \, , \quad i\in\{1,\ldots,n\} \, ,
\end{equation}
where $e$ is the empty word. Furthermore, we assume that
\begin{subequations}\label{eqn:series}
	\begin{align}
		S_i &= c + \sum_{j=1}^n S_{ij}x_j + \ldots\, ,\\
		S_i' &= c + \sum_{j=1}^n S_{ij}'x_j + \ldots\, ,
	\end{align}
\end{subequations}
with $\left(S_{ij}\right)_{i,j=1,\ldots,n}$ being invertible. Then there exists a unique homomorphism $\rho:\alg{f}(X)\,{\rightarrow}\, \alg{f}(X)$ such that
\begin{equation}\label{eqn:reqmap}
	\rho(S_i)=S_i' \, , \quad i\in\{1,\ldots,n\} \, .
\end{equation}
Let us now prove this statement. We make an ansatz for the map $\rho$ by writing
\begin{equation}
	\rho(x_j) = \sum_{w\in X^\times}\rho_w(x_j)\, w \, , \quad j\in\{1,\ldots,n\} \, .
\end{equation}
We now construct the coefficients $\rho_w(x_j)$ inductively on the length of words in $X^\times$ based on the requirement \eqref{eqn:reqmap} up to a given length. Due to the assumption \eqref{eqn:constterm}, we can set $\rho_e(x_j)\texteq0$ for all $j\,{\in}\,\{1,\ldots,n\}$. For illustrative purposes, let us explicitly also do the case of length one. Given the previous considerations, we can write up to length one
\begin{equation}
	\rho(S_i) - S_i' = \sum_{k=1}^n\left[\sum_{j=1}^nS_{ij}\,\rho_{x_k}(x_j)-S_{ik}'\right]x_k + \ldots \, ,
\end{equation}
which is what we require to vanish. Since the matrix $(S_{ij})_{i,j=1,\ldots,n}$ is invertible, this immediately translates to 
\begin{equation}
	\rho_{x_k}(x_j) = \sum_{l=1}^n(S^{-1})_{jl}S'_{lk} \, .
\end{equation}
Let us now assume that we have constructed the map $\rho(x_j)$, $j\,{\in}\,\{1,\ldots,n\}$, up to length $n$ such that $c_w(\rho(S_i))\texteq c_w(S_i')$ for all words $w\,{\in}\, X^\times$ with length at most $n\,{\geq}\,1$. Up to length $n\,{+}\,1$, we can write
\begin{equation}
	\begin{aligned}
		\rho(S_i) - S_i' =& \sum_{\substack{w\in X^\times\\|w| = n+1}}\left[\sum_{v\in X^\times}c_v(S_i)\rho_{w}(v)-c_w(S_i')\right]w + \ldots,
	\end{aligned}
\end{equation}
where we have defined $|w|$ as the length of the word $w\,{\in}\, X^\times$ and used the induction hypothesis, which is why we are only left with words of length equal to $n+1$. Again, we require this to vanish for all $i\,{\in}\,\{1,\ldots,n\}$ and $w\,{\in}\, X^\times$ with $|w|\texteq n\,{+}\,1$. Let us (generically) take such a word $w$. Then, since $\rho_w(v)\texteq0$ for $|v|\,{>}\,|w|$, requiring the above coefficients to vanish is equivalent to the system of equations
\begin{equation}\label{eqn:cond}
	\sum_{j=1}^nS_{ij}\rho_w(x_j) = c_w(S_i') - \sum_{\substack{v\in X^\times\\2\leq |v| \leq n+1}}c_v(S_i)\rho_w(v)  \, .
\end{equation}
Notice that $\rho_w(v)$ is a known quantity for $v\,{\in}\, X^\times$ such that $2\,{\leq}\,|v|\,{\leq}\, n\,{+}\,1$ by the induction hypothesis and the fact that $\rho$ is a homomorphism. We can now conclude the proof by observing that \eqn{eqn:cond} can be solved since we assume that the matrix $(S_{ij})_{i,j=1,\ldots,n}$ is invertible. Uniqueness is clear from the construction.

Beyond that, given that the series $S_i$ and $S_i'$ are group-like with respect to the canonical Hopf algebra structure on $\alg{f}(X)$ for all $i\,{\in}\,\{1,\ldots,n\}$, it immediately follows that $\rho$ commutes with the respective coproduct $\Delta$ (cf.~\eqn{eqn:coproduct}). In particular, the elements $\rho(x_i)$ are primitive for all $i\,{\in}\,\{1,\ldots,n\}$. To see this, notice that we can write
\begin{equation}
	\Delta(\rho(S_i))=\Delta(S_i')=S_i'\otimes S_i' = \rho(S_i)\otimes \rho(S_i) = (\rho\otimes\rho)\Delta(S_i)
\end{equation}
for all $i\,{\in}\,\{1,\ldots,n\}$. Upon expansion, this directly implies that
\begin{equation}
	\sum_{j=1}^nS_{ij}\Delta(\rho(x_j))=\sum_{j=1}^nS_{ij}(\rho\otimes\rho)\Delta(x_j) \, ,
\end{equation}
which implies that $\rho$ commutes with $\Delta$ on all the generators since $(S_{ij})_{i,j=1,\ldots,n}$ is invertible. Hence we can conclude that $\rho$ commutes with $\Delta$ as it does so on the generators. This now also implies that $\rho(x_i)$ is primitive as $x_i$ is primitive for $\Delta$ by definition. 

In this work, this theorem can be applied in the case $n\texteq2$, $\alg{f}(X)\texteq\alg{f}_2(\ellalph)$ for the power series $S_1 \texteq \reverse(\overline{A(\tau)})$, $S_1'\texteq A^{-1}(\tau)$, $S_2 \texteq \reverse(\overline{B(\tau)})$ and $S_2'\texteq B^{-1}(\tau)$ in order to solve the elliptic single-valued conditions \eqref{eqn:svgenusone} and prove the shuffle relations for the sveMPLs.

\appendixsection{Explicit expansions}

\appendixsubsection{\texorpdfstring{Expansions of $a'$ and $b'$}{Expansions of a' and b'}}\label{app:apbp}
The expansions of the letters of the second alphabet $\ellalph'\texteq\{a',b'\}$ in terms of the alphabet $\ellalph\texteq\{a,b\}$ after solving the elliptic single-valued conditions \eqref{eqn:svgenusone} and using the reversal map $\reverse(a)\texteq a,\,\reverse(b)\texteq b$, reads up to word length three
\begin{subequations}
\begin{align}
	a' &= -a - \left(2\Re(\coeffA(0,1|\tau))+\frac{\pi}{3}\Im(\tau)\right)ba^2 
	+ \left(4 \Re(\coeffA(0,1|\tau))+\frac{2\pi}{3}\Im(\tau)\right)aba\notag\\
	&\quad-\left(2 \Re(\coeffA(0,1|\tau))+\frac{\pi}{3}\Im(\tau)\right)a^2b+\cO(|w|\,{\geq}\,4)\notag\\
	&= -a - \frac{2}{\pi}\left(\zeta_2\Im(\tau)+\pi\Re(\coeffA(0,1|\tau))\right)[[b,a],a]+\cO(|w|\,{\geq}\,4),\\
	b' &= -\frac{\Im(\tau)}{\pi}a +b
	-\frac{2}{\pi}\left(\zeta_2\Im(\tau)+\pi\Re(\coeffA(0,1|\tau))\right)b^2a+\frac{4}{\pi}\left(\zeta_2\Im(\tau)+\pi\Re(\coeffA(0,1|\tau))\right)bab\notag\\
	&\quad-\frac{2}{\pi}\left(\zeta_2\Im(\tau)+\pi\Re(\coeffA(0,1|\tau))\right)ab^2-\frac{2\Im(\tau)}{\pi^2}\left(\zeta_2\Im(\tau)+\pi\Re(\coeffA(0,1|\tau))\right)ba^2\notag\\
	&\quad+\frac{4\Im(\tau)}{\pi^2}\left(\zeta_2\Im(\tau)+\pi\Re(\coeffA(0,1|\tau))\right)aba-\frac{2\Im(\tau)}{\pi^2}\left(\zeta_2\Im(\tau)+\pi\Re(\coeffA(0,1|\tau))\right)a^2b\notag\\
	&\quad+\cO(|w|\,{\geq}\,4)\notag\\
	&= -\frac{\Im(\tau)}{\pi}a +b
	-\frac{2}{\pi}\left(\zeta_2\Im(\tau)+\pi\Re(\coeffA(0,1|\tau))\right)\left([b,[b,a]]+\frac{\Im(\tau)}{\pi}[[b,a],a]\right)\notag\\
	&\quad+\cO(|w|\,{\geq}\,4).,
\end{align}
\end{subequations}
where the relation \eqref{eqn:modrelFinal} shown in \appref{app:coeffid} was used here to arrive at these simple expressions.

\appendixsubsection{\texorpdfstring{Expansion of $\mathbf{\Gamma}(z|\tau)$}{Expansion of Γ(z|τ)}}\label{app:svexp}
Up to second order in the words of letters $a$ and $b$, we can calculate the generating function of sveMPLs $\mathbf{\Gamma}(z|\tau)$ as
\begin{align}\label{eqn:exp}
	\mathbf{\Gamma}(z|\tau)=1&+\left(2\iunit\Im\left(\emplcoeff{a}{z}\right)-\frac{\Im(\tau)}{\pi}\overline{\emplcoeff{b}{z}}\right)a+2\Re\left(\emplcoeff{b}{z}\right)b\notag\\
	&+\Bigg(2\Re\left(\emplcoeff{a^2}{z}\right)+\frac{\Im(\tau)}{\pi}\left(\overline{\emplcoeff{ab}{z}}+\overline{\emplcoeff{ba}{z}}\right)-\emplcoeff{a}{z}\overline{\emplcoeff{a}{z}} \notag\\
	&\hspace{32ex}-\frac{\Im(\tau)}{\pi}\emplcoeff{a}{z}\overline{\emplcoeff{b}{z}}+\frac{\Im(\tau)^2}{\pi^2}\overline{\emplcoeff{b^2}{z}}\Bigg)a^2\notag\\
	&+\left(2\Re\left(\emplcoeff{b^2}{z}\right)+\emplcoeff{b}{z}\overline{\emplcoeff{b}{z}}\right)b^2 \notag\\
	&+\left(\emplcoeff{ab}{z}-\overline{\emplcoeff{ba}{z}}+\emplcoeff{a}{z}\overline{\emplcoeff{b}{z}}-\frac{\Im(\tau)}{\pi}\overline{\emplcoeff{b^2}{z}}\right)ab \notag\\
	&+\Bigg(\emplcoeff{ba}{z}-\overline{\emplcoeff{ab}{z}}-\overline{\emplcoeff{a}{z}}\emplcoeff{b}{z} \notag\\
	&\hspace{32ex}-\frac{\Im(\tau)}{\pi}\emplcoeff{b}{z}\overline{\emplcoeff{b}{z}} -\frac{\Im(\tau)}{\pi}\overline{\emplcoeff{b^2}{z}}\Bigg)ba \notag\\
	&+ \cO(|w|\geq 3)\,.
\end{align}
In certain cases, we can explicitly calculate the appearing integrals, e.g.
\begin{equation}\label{eqn:explicitInts}
	\begin{aligned}
		\emplcoeff{a^n}{z}&=(-1)^n\empltg{0^n}{z}=\frac{(-z)^n}{n!}\, ,\\
		\emplcoeff{b^n}{z}&=(-1)^n\int_0^z(-\nu(t_1|\tau))\int_0^{t_1}\cdots\int_0^{t_{n-1}}(-\nu(t_n|\tau))=\left(\frac{2\pi\iunit}{\Im(\tau)}\right)^n\frac{\Im(z)^n}{n!}\, ,\\
		\emplcoeff{ab}{z}&=\empltg{1}{z} \, ,\\
		\emplcoeff{ba}{z}&=-\empltg{1}{z}-2\pi\iunit z \frac{\Im(z)}{\Im(\tau)} \, ,
	\end{aligned}
\end{equation}
where we have used \eqn{eqn:nu} as well as the shuffle relations. Plugging this into \eqn{eqn:exp}, we can simplify the expansion to
\begin{equation}
	\mathbf{\Gamma}(z|\tau)=1+\left(-2\Re\left(\empltg{1}{z}\right)+2\pi\frac{\Im(z)^2}{\Im(\tau)}\right)[b,a] + \cO(|w|\geq 3) \, .
\end{equation}

\newpage


\bibliographystyle{esv}
\bibliography{esv}

\end{document}